\documentclass[aps,prd,showpacs,onecolumn,preprintnumbers,notitlepage,nofootinbib,tightenlines,11pt]{revtex4-1}
\usepackage{amsmath,bm}
\usepackage{times}
\usepackage{dcolumn}
\usepackage{amsfonts}
\usepackage{amssymb}
\usepackage{braket}
\usepackage{color,graphicx}
\usepackage{epstopdf}
\usepackage{latexsym}
\usepackage[dvipsnames]{xcolor}
\usepackage{slashed}
\usepackage{CJK,upgreek,fancyhdr}
\usepackage{float}
\usepackage{multirow}
\usepackage{hyperref}
\hypersetup{colorlinks,citecolor=nicegreen,linkcolor=nicered,urlcolor=RoyalBlue}
\usepackage{enumitem}
\usepackage{natbib}
\usepackage{relsize}
\usepackage[left=2.5cm,right=2.5cm,top=2.0cm,bottom=2.5cm]{geometry}
\newcommand{\beq}{\begin{eqnarray}}
\newcommand{\eeq}{\end{eqnarray}}
\newcommand{\non}{\nonumber\\ }

\newcommand{\psl}{ P \hspace{-2.4truemm}/ }
\newcommand{\nsl}{ n \hspace{-2.2truemm}/ }
\newcommand{\vsl}{ v \hspace{-2.2truemm}/ }
\newcommand{\epsl}{\epsilon \hspace{-1.6truemm}/\,  }

\def\lsim{ {\ \lower-1.2pt\vbox{\hbox{\rlap{$<$}\lower6pt\vbox{\hbox{$\sim$}
		}}}\ } }
\def\gsim{ {\ \lower-1.2pt\vbox{\hbox{\rlap{$>$}\lower6pt\vbox{\hbox{$\sim$}
		}}}\ } }

\def \jhep{ J. High Energy Phys.  }

\definecolor{Red}{rgb}{1.,0.,0.}
\definecolor{Blue}{rgb}{0.,0.,1.}
\definecolor{RoyalBlue}{rgb}{0.0,0.14,0.4}
\definecolor{nicered}{rgb}{0.7,0.1,0.2}
\definecolor{nicegreen}{rgb}{0.1,0.4,0.2}

\newcommand{\Gre}[1]{{\color{nicegreen}{#1}}}
\newcommand{\RyB}[1]{{\color{RoyalBlue}{#1}}}
\def\orcid#1{\kern .08em\href{https://orcid.org/#1}{\includegraphics[keepaspectratio,width=0.76em]{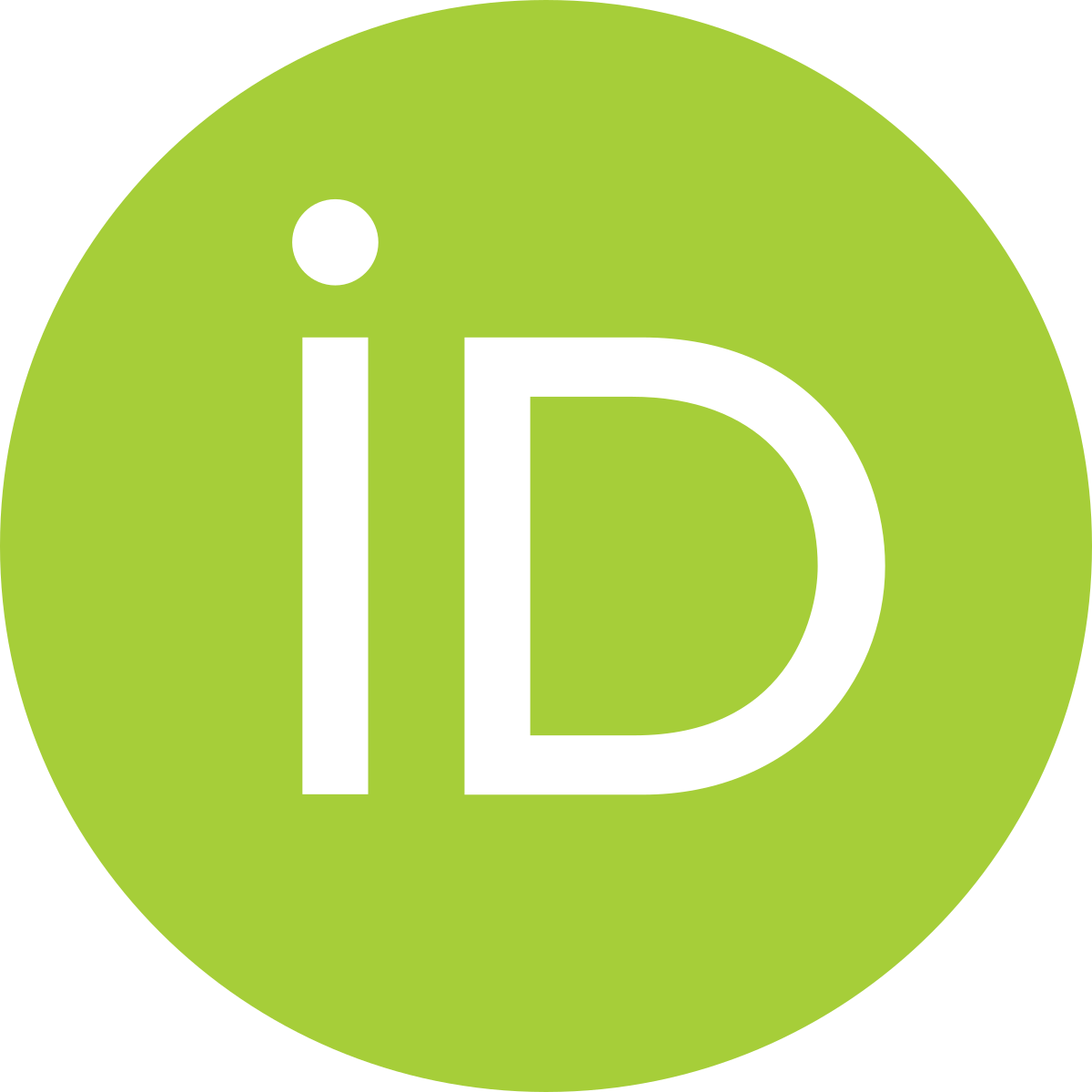}}}

\begin{document}
\title{\boldmath  The $B^0 \to J/\psi f_0(1370,1500,1710)$ decays:
an opportunity for scalar glueball hunting
}
\author{Jia-Le~Ren\orcid{0000-0001-8661-297X}}
\author{Min-Qi~Li}
\author{Xin~Liu\orcid{0000-0001-9419-7462}}
\affiliation{Department of Physics,
Jiangsu Normal University, Xuzhou 221116, China}
	
\author{Zhi-Tian~Zou}
\author{Ying~Li\orcid{0000-0002-1337-7662}}
\affiliation{ Department of Physics, Yantai University, Yantai 264005,China}
	
\author{Zhen-Jun~Xiao\orcid{0000-0002-4879-209X}}
\affiliation{Department of Physics and Institute of Theoretical Physics,\\
Nanjing Normal University, Nanjing 210023, China}

\date{\today{}}
	
\begin{abstract}
		
The scalars $f_0$  closest to 1.5~GeV contain the mesons $f_0(1370)$, $f_0(1500)$ and $f_0(1710)$,
and the latter two ones are usually viewed as the potential candidates for the scalar glueballs.
In this work, by including the important contributions from the vertex corrections,
we study the decays $B^0 \to J/\psi f_0$ within the improved perturbative QCD
approach and analyze the possible scalar glueball hunting. Together with the two mixing
models, namely, $f_0(1500) (f_0(1710))$ being the primary scalar glueball in model I (II),
and two classification scenarios, namely, $f_0$ being the $q\bar q$ excited (ground)
states in scenario 1 (2), the branching fractions associated with their ratios
for $B^0 \to J/\psi f_0$ are evaluated comprehensively. The predictions with still large uncertainties
in the considered two mixing models are roughly consistent with currently limited data,
which indicates that both more rich data and more precise predictions
are urgently demanded to figure out the scalar glueball clearly in the future.
Moreover, several interesting ratios between the branching fractions
of $B^0 \to J/\psi f_0(\to \pi^+ \pi^-/K^+ K^-)$ and $B^0 \to J/\psi \rho^0/\phi
(\to \pi^+ \pi^-/ K^+ K^-)$ that could help us to understand the nature of scalar $f_0$
are defined and predicted theoretically. These ratios should be examined in
future experiments.

\end{abstract}
	
\preprint{\footnotesize  JSNU-PHY-HEP-02/23}
\maketitle

	
\newpage
%
%
\section{Introduction}
\label{sect:1}
	
Quantum ChromoDynamics (QCD)~\cite{Fritzsch:1972jv,Gross:2022hyw}, a theory simultaneously involving
perturbative asymptotic freedom and nonperturbative color confinement, predicts firmly
the existence of gluonic bound states while without any constituent quarks~\cite{Fritzsch:1975tx},
namely, the glueballs. However, the definite identification of the glueballs has been proven challenging
and still remains a longstanding problem in hadron physics. Nevertheless, tremendous
efforts from both the theoretical and experimental sides have been devoted
to this subject. Some comprehensive reviews on the current status of glueballs could be
seen in, e.g., Refs.~\cite{Klempt:2007cp,Ochs:2013gi,Chen:2022asf}, and more references therein.

So far, based on Lattice QCD (LQCD) simulations, it is commonly believed that the lightest
glueball state should be a scalar with quantum number $J^{PC}=0^{++}$ and with mass around
1.5-1.8 GeV~\cite{Bali:1993fb,Chen:1994uw,Morningstar:1999rf,Vaccarino:1999ku,Lee:1999kv,Liu:2000ce,
Liu:2001je,Ishii:2001zq,Chen:2005mg,Loan:2005ff,Richards:2010ck,Gregory:2012hu,Gui:2012gx}.
While, three scalar resonances closet to this mass region, i.e., $f_0(1370)$, $f_0(1500)$
and $f_0(1710)$, could be found in the particle list provided by
the Particle Data Group (PDG)~\cite{ParticleDataGroup:2020ssz}
( Henceforth, we will adopt $f_0$ to generally denote these three scalars, unless otherwise
stated.). But, only one of them would be the potential candidate, that is, primarily a scalar
glueball~\footnote{\RyB{ In very recent studies~\cite{Klempt:2021nuf}, a distinct viewpoint,
that is, the concept of fragmented scalar glueball rather than the primary scalar glueball,
was proposed. Undoubtedly, this proposal makes the identification of scalar glueball
controversial but more interesting. This issue will be left for future investigations. }}.
It is due to the fact that the lowest-lying scalar glueball has the same
quantum number as QCD vacuum, and thereby mixes with scalar quarkonium~\cite{Lu:2013jj}.
Hence, the $q\bar{q}$ assignment cannot accommodate all these three $f_0$
states~\cite{Cheng:2006hu,Cheng:2015iaa}.

Accompanied with the discovery of
$f_0(1370)$ and $f_0(1500)$~\cite{CrystalBarrel:1994doj,Amsler:1995gf,CrystalBarrel:1996wfh}, 	
Amsler and Close proposed initially a flavor mixing scheme with the scalar quarkonia
$N (\equiv (u\bar u +d \bar d)/\sqrt{2})$ and $S (\equiv s\bar s$) and the scalar glueball
$G$ to produce the observed $f_0(1370)$, $f_0(1500)$
and $f_0(1710)$~\cite{Amsler:1995tu,Amsler:1995td}.
And, two of them can be considered
as $q\bar q$ mesons with glue content, and the rest one would be primarily the scalar
glueball but with (a) few scalar quarkonia. The $f_0$ state can then be expressed
as follows,
\beq
\vert f_0 \rangle &=& \alpha_1 \vert N \rangle
+ \alpha_2 \vert S \rangle
+ \alpha_3 \vert G \rangle\;, \label{eq:eigenstate}
\eeq
with coefficients $\alpha_i(i=1,2,3)$ measuring different kinds of contents in
$f_0$, and satisfying approximately the following normalization~\cite{Lu:2013jj},
\beq
{\vert \alpha_1 \vert}^2 + {\vert \alpha_2 \vert}^2 + {\vert \alpha_3 \vert}^2 \simeq 1.
\label{eq:normalization}
\eeq
In the past decades, numerous
studies have been made with different mixing schemes
\cite{Lee:1999kv,Li:2000yn,Close:2000yk,Li:2000cj,Close:2001ga,Giacosa:2004ug,
Close:2005vf,Giacosa:2005zt,Giacosa:2005qr,Cheng:2006hu,Chatzis:2011qz,Gui:2012gx,Janowski:2013uga,Janowski:2014ppa,
Cheng:2015iaa,Frere:2015xxa,Vento:2015yja,Noshad:2018afw,Guo:2020akt}. But, most of these studies
focused on the productions of $f_0$ in $J/\psi$ and decay properties of $f_0$ in the low energy region.

\begin{figure}[!!htb]
\centering
\begin{tabular}{l}
\includegraphics[width=0.85\textwidth]{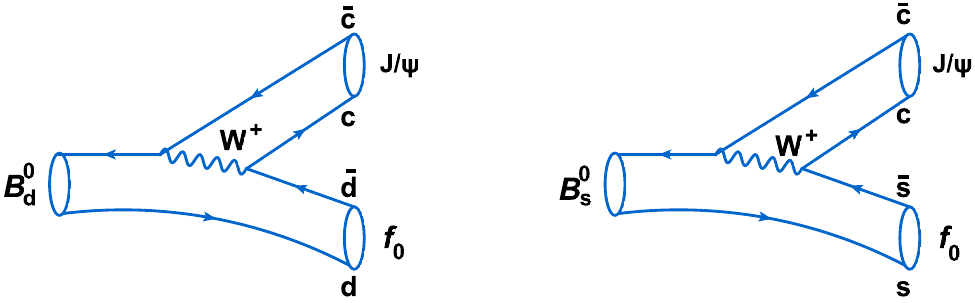}
\end{tabular}
\caption{(Color online) Leading quark-level Feynman diagrams for $B^0 \to J/\psi f_0$}
\label{fig:fig1}
\end{figure}

In contrast to the decays of $J/\psi$ or charmed mesons, the productions of $f_0$ in
$B$-meson (including doubly heavy $B_c$ meson) decays can also
provide
additional information,
because of the larger phase space and an
apparent suppression of higher spin ($J=2$) states~\cite{Minkowski:2004xf,He:2006qk,Chen:2007uq,Wang:2009rc,Wang:2009cb,Ochs:2013gi,Lu:2013jj,He:2015owa}.
Particularly, the experiments at $B$ factories, Large Hadron Collider, and Belle-II have
produced a large amount of events of $B$ decays, in which the scalar particles have been
detected ever since the first observation of $f_0(980)$ by Belle and {\it BABAR}
Collaborations~\cite{Abe:2002av,Aubert:2003mi}. Therefore, the study of $f_0$'s production
in $B$ decays could provide another effective way to clarify the intrinsic structure of scalars
and help to figure out the gluon component inside,
for example, the neutral $B$-meson decays into $f_0$ plus charmonia, in particular,
in light of the recent data on $B^0 \to J/\psi f_0(\to \pi^+ \pi^-/ K^+ K^-)$ decays~\cite{ParticleDataGroup:2020ssz,HFLAV:2022pwe}.
Here, $B^0$ includes $B_d^0$ and $B_s^0$.
The differentiation on flavor composition would be possible from the decays $B_d^0 \to
J/\psi f_0$ and $B_s^0 \to J/\psi f_0$ allowing for an isolation of
the scalar $N(d\bar d)$ and $S(s\bar s)$ components of $f_0$ correspondingly~\cite{Ochs:2013gi},
as illustrated in Fig.~\ref{fig:fig1}.
Therefore, the coefficients for related scalar quarkonia could be determined in $f_0$
through the $B^0 \to J/\psi f_0$ branching fractions (${\cal B}$), which may be useful for conjecturing
the fractions of scalar glueball~\footnote{\RyB{ In principle, when the precise distribution amplitudes of scalar
glueball are available, the studies of $B \to J/\psi G$ would be
possible based on factorization frameworks. It then looks more feasible to hunt for
the scalar glueball via $B$-meson decays directly. However, as claimed in Ref.~\cite{Lu:2013jj},
the form factor of $B$ to scalar glueball is suppressed by a factor
of 6-10 relative to that of $B$ to scalar mesons, then the small contributions from
$B$ to scalar glueball transition will be neglected tentatively in this work and are left for future studies,
though the color magnetic operator $O_{8g}$ in $B$-meson decays has a large Wilson
coefficient that could easily produce a number of gluons.}}.

Experimentally, the decays $B_s^0 \to J/\psi f_0(1370)$ and $B_s^0 \to J/\psi f_0(1500)$ have
been measured by Belle and Large Hadron Collider-beauty (LHCb) Collaborations through
$f_0 \to \pi^+ \pi^-$~\cite{Belle:2011phz,LHCb:2012ae,LHCb:2014ooi}.
Their branching fractions are reported as follows~\cite{ParticleDataGroup:2020ssz},
\beq
{\cal B}(B_s^0 \to J/\psi f_0(1370), f_0 \to \pi^+ \pi^-)
&=&
(0.34^{+0.14}_{-0.15}) \times 10^{-4}\;, \;\;\ ({\rm Belle})
\label{eq:psif13s-pp-ex-belle}\\
{\cal B}(B_s^0 \to J/\psi f_0(1370), f_0 \to \pi^+ \pi^-)
&=&
(4.4_{-4.0}^{+0.6})\times 10^{-5}
\;,\;\;\;\;\;\;({\rm LHCb})
\label{eq:psif13s-pp-ex-lhcb}
\\
{\cal B}(B_s^0 \to J/\psi f_0(1500), f_0 \to \pi^+ \pi^-)
&=&
(2.04_{-0.24}^{+0.32})\times 10^{-5}
\;.\;\;\;\;({\rm LHCb})
\label{eq:psif15s-pp-ex-lhcb}
\eeq
Of course, the large uncertainties are expected to be reduced by the
upgraded LHCb and the on-going Belle-II experiments.

Theoretically, the decays $B^0 \to J/\psi f_0$ have been investigated partially in Refs.~\cite{Xie:2014gla,Wang:2009rc,Wang:2009cb,Lu:2013jj}, in particular,
\begin{itemize}
\item
Under the approach based on chiral unity theory, the decays $B^0 \to J/\psi f_0(1370)$ and
$B^0 \to J/\psi f_0(1710)$ were studied with the assumptions of the scalar glueball $f_0(1500)$.
The ratios among their decay widths were also predicted with large uncertainties
as follows~\cite{Xie:2014gla},
\beq
\frac{\Gamma(B_d^0 \to J/\psi f_0(1370))}{\Gamma(B_d^0 \to J/\psi f_0(1710))} &=& 6.2 \pm 1.6\;,
\non
\frac{\Gamma(B_d^0 \to J/\psi f_0(1710))}{\Gamma(B_s^0 \to J/\psi f_0(1710))} &=& (7.7 \pm 1.9) \times 10^{-3}\;,
\label{eq:Rs-Xie}
\\
\frac{\Gamma(B_s^0 \to J/\psi f_0(1370))}{\Gamma(B_s^0 \to J/\psi f_0(1710))} &=& (1.1 \pm 0.3) \times 10^{-2}\;.
\nonumber
\eeq
		
\item
In Refs.~\cite{Wang:2009rc,Wang:2009cb}, using naive factorization and the Wilson coefficient
$a_2$ extracted from $B_d^0 \to J/\psi K_S^0$, the authors
estimated the branching fractions of $B_d^0 \to J/\psi f_0(N)$ and $B_s^0 \to J/\psi f_0(S)$
by classifying $f_0$ states in two scenarios as,
\beq
{\cal B}(B_d^0 \to J/\psi f_0(N))&\simeq& \left\{ \begin{array}{ll}
(23^{+12}_{-14})  \times 10^{-6}
& \vspace{0.12cm} \\
(10^{+7}_{-5})  \times 10^{-5}
&  \\ \end{array} \right. \;,
\label{eq:psif0-nn-NF}\\
{\cal B}(B_s^0 \to J/\psi f_0(S))&\simeq& \left\{ \begin{array}{ll}
(6.5^{+4.0}_{-4.5}) \times 10^{-4}
& \vspace{0.12cm} \\
(3.5^{+2.3}_{-1.4})  \times 10^{-3}
&  \\ \end{array} \right. \;.
\label{eq:psif0-ss-NF}
\eeq
The authors also claimed that such large branching fractions would offer an opportunity
to probe the structures of scalars and to solve the mixing problem between the scalar
mesons with the available data in the future, which can serve for inferring about
the glueball component in principle.

\item
The authors further focused on the potential identification
of a scalar glueball by extracting the coefficient $\alpha_1$ in $B_d^0 \to J/\psi f_0$
or $\alpha_2$ in $B_s^0 \to J/\psi f_0$ based on SU(3) flavor symmetry and naive
factorization, but the definite conclusion could not be drew~\cite{Lu:2013jj}.
\end{itemize}
	
In this work, we will study the decays $B^0 \to J/\psi f_0$ in a comprehensive manner
by employing the perturbative QCD(PQCD) approach~\cite{Keum:2000ph,Keum:2000wi,Lu:2000em,Lu:2000hj}
at the known next-to-leading order (NLO) accuracy
to analyze the opportunity for identifying the possible scalar glueball potentially.
Different from QCD factorization approach~\cite{Beneke:1999br,Du:2000ff}
and soft-collinear effective theory~\cite{Bauer:2000yr}
based on the collinear factorization theorem, the PQCD approach within the framework of
$k_T$ factorization theorem has advantages to deal with
the nonfactorizable emission ($nfe$) diagrams and the annihilation ones~\footnote{
\RyB { Very recently, our colleagues have improved the framework of QCD
factorization on perturbatively calculating the weak annihilation diagrams in $B$-meson
decays~\cite{Lu:2022kos}. Then the inevitable parameterizations due to
the unavoidable end-point divergences in this
approach would be a thing of the past
and the predictive power is therefore believed to be recovered gradually.}},
besides the factorizable emission ($fe$) contributions.
By keeping the
transverse momentum $k_T$ of quarks,
the PQCD approach avoid the end-point divergences.
Furthermore, with resummation techniques, two resultant factors are produced
to guarantee the removal of the end-point singularities, which makes
PQCD calculations of hadronic matrix elements effective and reliable.
One called Sudakov factor, $e^{-S(t)}$ with $t$ the running scale at the
largest energy, could strongly suppress the soft dynamics
via resumming the double logarithms in the small $k_T$ (or large ${\bf b}$,
the conjugate space coordinate of transverse momentum $k_T$) region with
$k_T$ resummation~\cite{Botts:1989nd,Li:1992nu}. And another one is
called jet function, $S_t(x)$ with $x$ the momentum fraction
of valence quark in a meson, which could largely
smear the end-point divergences through resumming the double logarithms
in the small $x$ region with threshold resummation~\cite{Li:2001ay,Li:2002mi}.
The detailed expressions of $S_t(x)$ and $e^{-S(t)}$ are
referred to the Refs.~\cite{Li:2001ay,Li:2002mi,Botts:1989nd,Li:1992nu,Li:2003yj,Liu:2018kuo}.
In recent years, several developments on the PQCD approach have been
obtained, for a review, see, e.g.~\cite{Cheng:2020fcx}.
Particularly, the newly derived Sudakov factor for $J/\psi$ by including the charm quark mass effects further improves
the PQCD framework for $B$-meson decaying into charmonia plus light hadron(s)~\cite{Liu:2018kuo,Liu:2023kxr}.

As presented in, e.g., Refs.~\cite{Cheng:2000kt,Song:2002gw,Chen:2005ht,Li:2006vq,Beneke:2008pi,
Liu:2009yno,Colangelo:2010wg,Liu:2012ib,Liu:2013nea,Wang:2015uea,Liu:2019ymi,Yao:2022zom},
the $B$-meson decays into a charmonium plus light hadrons
are color-suppressed and should include the NLO contributions
from vertex corrections and NLO Wilson coefficients to make predictions compatible
with data. Therefore, associated
with the newly derived Sudakov factor for charmonia~\cite{Liu:2018kuo,Liu:2023kxr},
we shall comprehensively evaluate the branching fractions and their interesting ratios
in $B^0 \to J/\psi f_0$ at the NLO accuracy, together with two available models for $f_0$'s mixing, i.e.,
$f_0(1500) (f_0(1710))$ is viewed as the primary scalar glueball in model I (II), and two possible
scenarios for $f_0$'s classification, i.e., $f_0$ is regarded as the two-quark first excited
(lowest-lying) mesons in scenario 1 (2). The predicted branching fractions could provide
a natural filter of quarkonia $N$ or $S$ and the scalar glueball
could be deduced from the ratios among those branching fractions.
On the other hand, analogous to the channel
$B_s^0 \to J/\psi f_0(980)$, the decays $B_s^0 \to J/\psi f_0$ with no needs of
angular analysis
could also contribute clearly to the CP-violating parameter, namely, the $B_s^0$-$\bar B_s^0$
mixing phase $\beta_s$, which is sensitive to the possible new physics beyond the standard model.

This paper is organized as follows. In Sect.~\ref{sect:2}, we give a brief review
on the $f_0$'s mixing and classifications that will be adopted.
Perturbative calculations of the $B^0 \to J/\psi f_0$
decay amplitudes in the PQCD approach are also collected in this section.
In Sect.~\ref{sect:3}, we perform the numerical evaluations and remark on the
theoretical results phenomenologically.
A summary of this work is finally given in Sect.~\ref{sect:4}.

%
%
\section{PERTURBATIVE QCD CALCULATIONs}
\label{sect:2}
	
\subsection{\boldmath Classification and flavor mixing of $f_0$} \label{ssec:2-1}

A well-known fact is that the inner structure of light scalars is not
yet well established theoretically (for a review, see, e.g.,
Refs.~\cite{Klempt:2007cp,Ochs:2013gi,Chen:2022asf}).
Many explanations to their possible contents
have been proposed, for example, $\bar qq$, $\bar q\bar qqq$, meson-meson bound states or even
supplemented with a scalar glueball. It seems that they are
not made of one simple component but are the superpositions of the above mentioned ones.
Actually, different scenarios tend to provide different predictions on the production and decay of light
scalars, which are helpful to determine the dominant component.

Nowadays, in the spectroscopy study, many light scalar states have been discovered
experimentally but their underlying structures are still remaining basically unknown. According to the
particles collected by PDG~\cite{ParticleDataGroup:2020ssz}, the light scalars below or near 1
GeV, including $a_0(980)$, $K^*_0(700)({\rm or}~ \kappa)$,
$f_0(500)({\rm or}~ \sigma)$, and $f_0(980)$, are usually viewed
to form an SU(3) flavor nonet; while the light scalars around 1.5 GeV,
including $a_0(1450)$, $K^*_0(1430)$, $f_0(1370)$, and
$f_0(1500)/f_0(1710)$, form another nonet.
	
Presently, it is generally accepted that the light scalars can be considered as $q\bar q$-mesons in
two scenarios~\cite{Cheng:2005nb}, namely,
\begin{itemize}
\item {Scenario 1 (S1):} the light scalars in the aforementioned former nonet are
treated as the lowest-lying $q\bar q$ states, and those in the latter
nonet are the corresponding first excited states.
		
\item {Scenario 2 (S2):} the light scalars in the aforementioned latter nonet are
viewed as the ground $q\bar q$ states and the corresponding first
excited ones are believed to lie between $(2.0\sim 2.3)$~GeV. While those in the former nonet
have to be four-quark bound states.
\end{itemize}
Therefore, in the two-quark picture, the scalars $f_0(1370)$ and $f_0(1500)/f_0(1710)$
considered in this work will be $q\bar q$ nonet corresponding to the first excited states
in S1 while the lowest-lying states in S2.

Now, let us briefly review the status about the scalar quarkonia and glueball mixing
schemes, namely, $N$-$S$-$G$ mixing, for the scalars $f_0(1370)$, $f_0(1500)$, and $f_0(1710)$. 	
As aforementioned, ever since the pioneering works on the mixing
of the scalar quarkonia and the scalar glueball by Amsler and Close, a number
of schemes have been proposed
generally by combining the experimental data
and LQCD calculations.
A consensus has been reached that $f_0(1370)$ does not have a sizable $G$ component
and is predominated by the scalar quarkonium $N$, which is consistent with the fact that
the $f_0(1370)$ does not couple strongly to $K\bar{K}$ ~\cite{Fritzsch:1972jv}, though
it is quite controversial that which of the two remaining isoscalars, i.e., $f_0(1500)$ and $f_0(1710)$, is primarily the scalar glueball.
	
Among all the available mixing schemes
in the literature, two different models
are proposed to describe the $N$-$S$-$G$ mixing. That is,
$f_0(1500)$ is primarily a scalar glueball while
$f_0(1710)$ is governed by the scalar quarkonium $S$ in
model  I ($M_{\rm I}$),
and, $f_0(1710)$ is prominently a scalar glueball while $f_0(1500)$ is dominated
by the scalar quarkonium $S$ in  model
II ($M_{\rm {II}}$). By fitting the masses, namely,
the scalar $N$-quarkonium mass $m_N$, the scalar $S$-quarkonium mass $m_S$
and the scalar glueball mass $m_G$, and analyzing their branching fractions
in strong decays, the matrix elements of $f_0(1370)$-$f_0(1500)$-$f_0(1710)$ mixing are constrained.
But, the magnitude and sign of the mixing matrix elements are usually different even if in the same kind of model.
It seems that the current status of the mixing schemes is highly complicated and evidently far from satisfactory,
implying that the definite determination of the matrices for $N$-$S$-$G$ mixing is still a tough task.
	
Despite of all that, due to no interferences between $B_d^0 \to J/\psi f_0(N)$ and $B_s^0 \to J/\psi f_0(S)$,
the sign in the mixing matrices does not affect the analyses in this work and will be left for future
studies involving significant interferences. Then we just take the magnitude of the mixing matrices
into account. To date, the mixing matrices have been studied mostly by following
various measurements and LQCD calculations with good precision. Therefore, in light of
the increasing accuracy of experimental measurements and LQCD evaluations, it is much
better for us to consider the mixing schemes in the literature since 2000.
	
For the sake of objectiveness in the following calculations and analyses, averaging the matrix elements~\cite{Hsiao:2014dva}
collected from various works~\footnote{\RyB{
It is worth emphasizing that, due to the limited space, we will quote the related matrices straightforwardly here but not
provide the unnecessary comments on why and how to obtain the matrix elements. The readers
could refer to the original references cited in this work. }} in each kind of model is preferred.
Explicitly, the averaged matrix elements $\bar \alpha_i(i=1,2,3)$ for the
scalar quarkonia $N$ and $S$ and the scalar glueball $G$ can be defined as follows,
\beq
\bar \alpha_i  & =&  \Biggl( \sum_{n} {|\alpha_i|}_n \Biggr)/ n \;,
\label{eq:aver-M}
\eeq
where $n$ is the number of mixing matrices quoted in this study.
Notice that, only the central values of matrix elements $\alpha_i$ will be quoted here for convenience.
In order to estimate the theoretical errors induced by the mixing matrix elements,
the uncertainties $\Delta \bar\alpha_i$ are given by varying
the averaged central values $\bar \alpha_i$ with ten percent. Therefore,
\begin{itemize}
\item 		
For model
I, by combining five mixing matrices
in Refs.~\cite{Close:2000yk,Close:2001ga,Giacosa:2004ug,Giacosa:2005qr,Chatzis:2011qz},
the averaged mixing matrix with $f_0(1500)$ being the scalar glueball is obtained as follows,
\beq
\left(\begin{array}{cccc}
\vert f_0(1370)\rangle \\
\vert f_0(1500)\rangle \\
\vert f_0(1710)\rangle \\
\end{array} \right )
=\left(\begin{array}{cccc}
0.810\pm 0.081 & 0.164\pm 0.016 & 0.540\pm 0.054\\
0.545\pm 0.055 & 0.230\pm 0.023 & 0.787\pm 0.079\\
0.169\pm 0.017 & 0.942\pm 0.094 & 0.229\pm 0.023\\
\end{array} \right )\
\left(\begin{array}{cccc}
\vert N
\rangle \\
\vert S
\rangle \\
\vert G \rangle \\
\end{array} \right )
\label{eq:MI}
\eeq
with the mass ordering $m_S > m_G > m_N$ and $m_G \simeq 1500$~MeV. 		
		
\item 		
For model
II, by including five mixing matrices in Refs.~\cite{Li:2000yn,Chatzis:2011qz,Janowski:2013uga,Janowski:2014ppa,Guo:2020akt},
the averaged mixing matrix with $f_0(1710)$ being the scalar glueball can be read as follows,
\beq
\left(\begin{array}{cccc}
\vert f_0(1370)\rangle \\
\vert f_0(1500)\rangle \\
\vert f_0(1710)\rangle \\
\end{array} \right )
=\left(\begin{array}{cccc}
0.862\pm 0.086 & 0.331\pm 0.033 & 0.311\pm 0.031\\
0.404\pm 0.040 & 0.869\pm 0.087 & 0.173\pm 0.017\\
0.247\pm 0.025 & 0.275\pm 0.028 & 0.912\pm 0.091\\
\end{array} \right )\
\left(\begin{array}{cccc}
\vert N
\rangle \\
\vert S
\rangle \\
\vert G \rangle \\
\end{array} \right )
\label{eq:MII}
\eeq
with the mass ordering $m_G > m_S > m_N$ and $m_G \simeq 1700$~MeV.
\end{itemize}
With them, we could then calculate the $B^0 \to J/\psi f_0$ branching fractions
in both models
I and II to analyze the possibility for potential scalar glueball hunting.
The explicit expressions and the information of citations about the quoted mixing matrices have been
collected in Appendix~\ref{sec:app1}.

\subsection{\boldmath{PQCD calculations of $B^0 \to J/\psi f_0$}} \label{ssec:2-2}
	
In the past two decades, PQCD approach, one of the popular factorization
approaches on the basis of QCD, has been widely employed to study varieties of $B$-meson decays.
Furthermore, the (quasi-) two-body $B$-meson decays into $J/\psi$ plus a light meson (or resonance)
~\cite{Chen:2005ht,Li:2006vq,Liu:2009yno,Liu:2012ib,Liu:2013nea,Wang:2015uea,Liu:2019ymi,Yao:2022zom}
have been studied in PQCD approach at the known NLO accuracy and most
theoretical predictions are in agreement with the current data.
Particularly, the PQCD study of the decays $B^0 \to J/\psi f_0(500,980)$~\cite{Liu:2019ymi}
provided basically consistent predictions
with the experimental measurements by different collaborations such as CDF, CMS, D0 and LHCb.
Therefore, for investigating the decays $B^0 \to J/\psi f_0(1370,1500,1710)$, it is natural
to follow the same analytic calculations presented in Ref.~\cite{Liu:2019ymi}.

Because of the massive beauty quark, relative to the light $u,d,$ and $s$ quarks, we will work in the $B^0$-meson rest frame for convenience.
The momenta $P_1$, $P_2$, and $P_3$ of the considered initial and final $B^0$, $J/\psi$, and $f_0$ mesons in the light-cone coordinates could be correspondingly written as,
\beq
P_{1}&=&\frac{m_{B}}{\sqrt{2}}(1, 1, {\bf 0}_{T})\;,
\qquad
P_{2}=\frac{m_{B}}{\sqrt{2}}(1-r_{3}^{2}, r_{2}^{2}, {\bf 0}_{T})\;,
\qquad
P_{3}=\frac{m_{B}}{\sqrt{2}}(r_{3}^{2}, 1-r_{2}^{2}, {\bf 0}_{T})\;,
\label{eq:momenta}
\eeq
where the $J/\psi (f_0)$ meson is chosen to move on the plus (minus) $z$-direction, $r_2 = m_{J/\psi}/m_{B^0}$, and $r_3 = m_{f_0}/ m_{B^0}$. The longitudinal polarization vector of $J/\psi$ is  $\epsilon_L = \frac{1}{\sqrt{2(1-r_3^2)} r_2}(1-r_3^2, -r_2^2, 0_{\bf T})$. Putting the (light) quark momenta in $B^0$, $J/\psi$, and $f_0$ mesons as $k_1$, $k_2$, and $k_3$, respectively, we have,
\beq
k_1 &=& (x_1 P_1^+, 0, k_{1 {\bf T}}), \qquad
k_2 = (x_2 P_2^+. x_2 P_2^-, k_{2{\bf T}}), \qquad
k_3 = (x_3 P_3^+, x_3 P_3^-, k_{3{\bf T}}).
\eeq
Then the $B^0 \to J/\psi f_0$ decay amplitude could be written in a factorized form,
\beq
{\cal A}(B^0 \to J/\psi f_0) &\sim &\int\!\! d x_1 d
x_2 d x_3 b_1 d b_1 b_2 d b_2 b_3 d b_3
\non &\times & {\rm Tr}
\left [ C(t) \Phi_{B}(x_1, b_1) \Phi_{J/\psi}(x_2, b_2)
\Phi_{f_0}(x_3, b_3) H(x_i, b_i, t) S_t(x_i)\, e^{-S(t)} \right ]\;.
\label{eq:DecAmp}
\eeq
in which, $x_{i}$ denotes the fraction of momentum carried by (light) quark in each meson,
$b_i$ is the conjugate space coordinate of $k_{i{\bf T}}$,  $C(t)$ stands for the Wilson
coefficients at scale $t$ with $t$ being the largest energy scale in the hard kernel
$H(x_i, b_i, t)$, and $\Phi$ is the wave function. The jet function $S_t(x_i)$ and the
Sudakov factor $e^{-S(t)}$ ensure the reliability of the perturbative calculations in
PQCD approach by effectively eliminating the end-point singularities and suppressing
the long-distance contributions~\cite{Keum:2000ph,Keum:2000wi,Lu:2000em,Lu:2000hj}.

For the $B^0$ meson, the light-cone wave function in the conjugate ${\bf b}$ space of
transverse momentum ${\bf k}_T$ can generally be defined
as~\cite{Keum:2000ph,Keum:2000wi,Lu:2000em,Lu:2000hj},
\beq
\Phi_{B^0}(x,{\bf b}) &=& \frac{i }{\sqrt{2N_c}} \biggl\{(\psl+m_{B^0}) \gamma_5
\phi_{B^0}(x, {\bf b}) \biggr\}_{\alpha\beta},
\label{eq:wf-B}
\eeq
where $\alpha$ and $\beta$ are the color indices and $N_c$ is the color factor.
The leading-twist distribution amplitude $\phi_{B^0}$ has the form widely
used in the PQCD approach as follows
\beq
\phi_{B^0}(x,{ b})&=& N_{B^0}x^2(1-x)^2
\exp\biggl[-\frac{1}{2}\left(\frac{x m_{B^0}}{\omega_{B^0}}\right)^2
-\frac{\omega_{B^0}^2 {b}^2}{2}\biggr] \;,
\eeq
where $\omega_{B^0}$ is the shape parameter of $\phi_{B^0}(x,{\bf b})$ and
$N_{B^0}$ is the normalization factor, satisfying the following normalization condition,
\beq
\int_0^1 dx \phi_{B^0}(x, b=0) &=& \frac{f_{B^0}}{2 \sqrt{2N_c}}\;.
\label{eq:norm}
\eeq
Note that, in principle, there are two Lorentz structures in $B^0$-meson
distribution amplitudes to be considered in the numerical calculations; however, the contribution induced
by the second Lorentz structure is numerically small and usually neglected~\cite{Lu:2000hj}.
In our calculations, the shape parameter $\omega_{B_{d (s)}^0} = 0.40 \pm 0.04 (0.50 \pm 0.05)$
GeV and the decay constant $f_{B_{d (s)}^0}=0.21 (0.23)$ GeV for $B_{d (s)}^0$ meson~\cite{Li:2005kt,Ali:2007ff}
are adopted. The recent developments on the $B$-meson distribution amplitude with high twists can be
found in~\cite{Bell:2013tfa,Feldmann:2014ika,Braun:2017liq,Wang:2019msf,Galda:2020epp}. The effects
induced by these newly developed distribution amplitudes will be left for future investigations together
with highly precise measurements.
		
For $J/\psi$ meson, the wave function in longitudinal polarization is presented as follows~\cite{Bondar:2004sv}
\beq
\Phi_{J/\psi}^{L}(x) &=& \frac{1}{\sqrt{2N_{c}}}
\biggl\{m \epsl_{L}\phi_{J/\psi}^{L}(x)
+\epsl_{L}\psl\ \phi_{J/\psi}^{t}(x)\biggl\}_{\alpha\beta}\;,
\label{eq:wf-psi-L}
\eeq
with $m, \epsilon_L$, and $P$ being the mass, the longitudinal polarization vector,
and the momentum of $J/\psi$, respectively. The twist-2 and -3 distribution amplitudes $\phi^{L}_{J/\psi}(x)$
and $\phi^{t}_{J/\psi}(x)$ are given as,
\beq
\phi_{J/\psi}^{L}(x) &=&
9.58\frac{f_{J/\psi}}{2\sqrt{2N_{c}}}x(1-x)
\biggl[\frac{x(1-x)}{1-2.8x(1-x)}\biggl]^{0.7}\;,
\label{eq:das-psi-LT}
\\
\phi_{J/\psi}^{t}(x) &=& 10.94\frac{f_{J/\psi}}{2\sqrt{2N_{c}}}(1-2x)^{2}\biggl[\frac{x(1-x)}{1-2.8x(1-x)}\biggl]^{0.7}\;.
\label{eq:da-psi-t}
\eeq
with $f_{J/\psi}$ being the $J/\psi$ decay constant.

The light-cone wave function for the scalar meson can be read as~\cite{Cheng:2005ye,Cheng:2005nb,Cheng:2013fba},
\beq
\Phi(x) &=& \frac{i}{\sqrt{2N_{c}}}
\biggl\{\psl\phi(x)+m \phi^S(x)+m (\nsl\vsl-1) \phi^T(x)\biggl\}_{\alpha\beta}\;,
\label{eq:wf-fq}
\eeq
with the twist-2 light-cone distribution amplitude (LCDA) $\phi(x)$ and
the twist-3 LCDAs
$\phi^{S, T}(x)$,
and with $m$ being the mass of scalar meson.
These LCDAs for the neutral scalar meson
can be expanded as Gegenbauer polynomials in the following forms,
\beq
\phi(x) &=& \frac{\bar{f}(\mu)}{2\sqrt{2N_{c}}}
\biggl\{6x(1-x)\sum_{n=1}^{\infty}B_n(\mu)C_n^{3/2}(2x-1)\biggl\}\;,
\label{eq:lcda-fq-t2}
\\ 
\phi^S(x) &=& \frac{\bar{f}(\mu)}{2\sqrt{2N_{c}}}
\biggl\{1+\sum_{n=1}^{\infty}a_n(\mu)C_n^{1/2}(2x-1)\biggl\}\;,
\label{eq:lcda-s-fq-t3g}
\\ 
\phi^T(x) &=& \frac{\bar{f}(\mu)}{2\sqrt{2N_{c}}}\frac{d}{dx}\biggl\{x(1-x)\times \biggl[1+\sum_{n=1}^{\infty}b_n(\mu)C_n^{3/2}(2x-1)\biggl]\biggl\},
\label{eq:lcda-t-fq-t3g}
\eeq
where $B_n(\mu)$, $a_n(\mu)$ and $b_n(\mu)$ are the Gegenbauer moments
and $C_n^{3/2}$ and $C_n^{1/2}$ are the Gegenbauer polynomials.
It is worth mentioning that, in contrast to the odd Gegenbauer moments vanishing for
the $\pi$ and $\rho$ mesons, the even Gegenbauer coefficients $B_n$ for the scalars are suppressed~\cite{Cheng:2005nb,Cheng:2013fba}.
Namely, the LCDA
$\phi(x)$ of the scalar meson is
governed by the odd Gegenbauer moments, in which, the $B_1$ and $B_3$ in S1 and S2 have been calculated by
using the QCD sum rule method~\cite{Cheng:2005nb}.
To date, the twist-3 LCDAs
with inclusion of Gegenbauer polynomials have been investigated, and the Gegenbauer moments
$a_{1,2,4}$ and $b_{1,2,4}$ have been calculated
only in S2~\cite{Lu:2006fr,Han:2013zg}. As implied in~\cite{Chen:2021dwn},
the twist-3 LCDAs contribute to the branching ratios slightly, while to the {\it CP}
violations significantly through annihilation diagrams.
The effects induced by the
Gegenbauer polynomials in the twist-3 distribution amplitudes
are hence left for future studies with more precise data.

For the decays $B^0 \to J/\psi f_0$, the effective Hamiltonian $H_{\rm eff}$ could be written
as~\cite{Buchalla:1995vs}
\beq
H_{\rm eff}\, &=&\, {G_F\over\sqrt{2}} \biggl\{ V^*_{cb}V_{cq} \biggl[ C_1(\mu)O_1^{c}(\mu)
+C_2(\mu)O_2^{c}(\mu) \biggr] - V^*_{tb}V_{tq} \biggl[ \sum_{i=3}^{10}C_i(\mu)O_i(\mu) \biggr] \biggr\}\;,
\label{eq:heff}
\eeq
where the light quark $q=d$ or $s$, the Fermi constant $G_F=1.16639\times 10^{-5}{\rm
GeV}^{-2}$. $V$ represents the Cabibbo-Kobayashi-Maskawa (CKM) matrix elements,
and $C_i(\mu)$ are Wilson coefficients at the renormalization
scale $\mu$. The local four-quark operators $O_i(i=1,\cdots,10)$ are listed in order:
\begin{enumerate}
\item[]{(1) Tree operators}
\begin{eqnarray}
{\renewcommand\arraystretch{1.5}
\begin{array}{ll}
\displaystyle
O_1^{c}\, =\,
(\bar{q}_\alpha c_\beta)_{V-A}(\bar{c}_\beta b_\alpha)_{V-A}\;,
& \displaystyle
O_2^{c}\, =\, (\bar{q}_\alpha c_\alpha)_{V-A}(\bar{c}_\beta b_\beta)_{V-A}\;,
\end{array}}
\label{eq:operators-1}
\end{eqnarray}
		
\item[]{(2) QCD penguin operators}
\begin{eqnarray}
{\renewcommand\arraystretch{1.5}
\begin{array}{ll}
\displaystyle
O_3\, =\, (\bar{q}_\alpha b_\alpha)_{V-A}\sum_{q'}(\bar{q}'_\beta q'_\beta)_{V-A}\;,
& \displaystyle
O_4\, =\, (\bar{q}_\alpha b_\beta)_{V-A}\sum_{q'}(\bar{q}'_\beta q'_\alpha)_{V-A}\;,
\\
\displaystyle
O_5\, =\, (\bar{q}_\alpha b_\alpha)_{V-A}\sum_{q'}(\bar{q}'_\beta q'_\beta)_{V+A}\;,
& \displaystyle
O_6\, =\, (\bar{q}_\alpha b_\beta)_{V-A}\sum_{q'}(\bar{q}'_\beta q'_\alpha)_{V+A}\;,
\end{array}}
\label{eq:operators-2}
\end{eqnarray}
		
\item[]{(3) Electroweak penguin operators}
\begin{eqnarray}
{\renewcommand\arraystretch{1.5}
\begin{array}{ll}
\displaystyle
O_7\, =\,
\frac{3}{2}(\bar{q}_\alpha b_\alpha)_{V-A}\sum_{q'}e_{q'}(\bar{q}'_\beta q'_\beta)_{V+A}\;,
& \displaystyle
O_8\, =\,
\frac{3}{2}(\bar{q}_\alpha b_\beta)_{V-A}\sum_{q'}e_{q'}(\bar{q}'_\beta q'_\alpha)_{V+A}\;,
\\
\displaystyle
O_9\, =\,
\frac{3}{2}(\bar{q}_\alpha b_\alpha)_{V-A}\sum_{q'}e_{q'}(\bar{q}'_\beta q'_\beta)_{V-A}\;,
& \displaystyle
O_{10}\, =\,
\frac{3}{2}(\bar{q}_\alpha b_\beta)_{V-A}\sum_{q'}e_{q'}(\bar{q}'_\beta q'_\alpha)_{V-A}\;,
\end{array}}
\label{eq:operators-3}
\end{eqnarray}
\end{enumerate}
in which, $\alpha$ and $\beta$ are the color indices and the notations
$(\bar{q}'q')_{V\pm A} = \bar q' \gamma_\mu (1\pm \gamma_5)q'$.
The index $q'$ in the summation of the above operators runs
through $u,\;d,\;s$, $c$, and $b$. Here, we define the Wilson coefficients as
\beq
a_{1}&=&C_{2}+\frac{C_{1}}{3}\;,
\qquad
a_{2}=C_{1}+\frac{C_{2}}{3}\;,
\\
a_{i} &=& C_{i}+\frac{C_{i + 1}}{3}(i=3,5,7,9)\;,
\qquad
a_j = C_j + \frac{C_{j -1}}{3}(j=4,6,8,10) \;.
\label{eq:wilson}
\eeq

\begin{figure}[htp]
\centering
\includegraphics[scale=0.9]{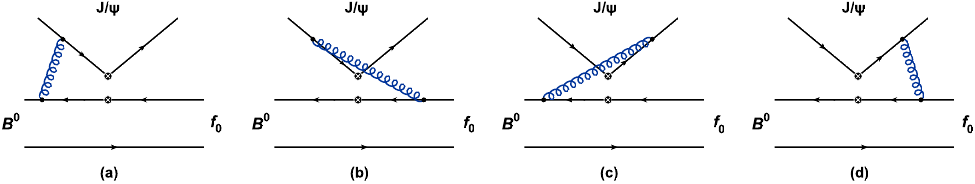}
\caption{(Color online) Vertex corrections to $B^0 \to J/\psi f_0$ }
\label{fig:fig2}
\end{figure}

\begin{figure}[htp]
\centering
\includegraphics[scale=1.0]{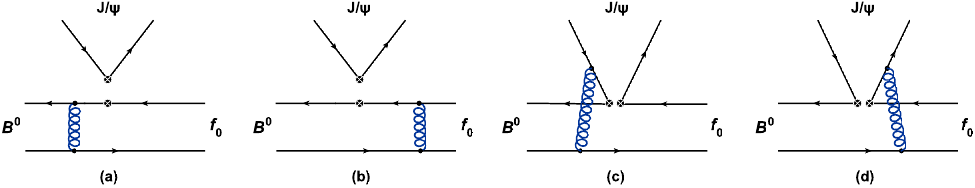}
\caption{ (Color online) Leading order Feynman diagrams for $B^0 \to J/\psi f_0$ in the PQCD approach
}
\label{fig:fig3}
\end{figure}

Before proceeding, two remarks are presented necessarily as follows,
\begin{itemize}	
\item		
In the literature, many works such as Refs.~\cite{Cheng:2000kt,Song:2002gw,Chen:2005ht,Li:2006vq,Beneke:2008pi,Liu:2009yno,Colangelo:2010wg,Liu:2012ib,
Liu:2013nea,Wang:2015uea,Liu:2019ymi,Yao:2022zom} have proved that the $B$-meson decays into
charmonia plus light hadrons, the color-suppressed decay modes, usually
receive the important NLO contributions, that is, vertex corrections.
Hence, for the decays $B^0 \to J/\psi f_0$,
the related vertex corrections depicted in Fig.~\ref{fig:fig2} will
contribute. As pointed out in~\cite{Cheng:2000kt}, their effects will modify the Wilson coefficients
in the factorizable emission diagrams Figs.~\ref{fig:fig3}(a) and \ref{fig:fig3}(b) and further lead to
a set of effective Wilson coefficients $\tilde{a}_i (i=2,3,5,7,9)$.
The detailed expressions of $\tilde{a}_i$ are given in~\cite{Liu:2019ymi}, and
no longer presented here.

\item
As stated in~\cite{Liu:2013nea}, when we consider the PQCD calculation at the NLO accuracy, it is
natural for us to include the NLO Wilson coefficients $C_{i}(m_{W})$ and the
NLO renormalization group evolution matrix $U(t,m,\alpha)$ for the Wilson coefficient
(see Eq.~(7.22) in~\cite{Buchalla:1995vs}) with the running coupling $\alpha_{s}(t)$ at two-loop,
\beq
\alpha_{s}(t)&=&\frac{4\pi}{\beta_{0}\ln(t^{2}/\Lambda_{\rm QCD}^{2})}\cdot \biggl
\{1-\frac{\beta_{1}}
{\beta_{0}^{2}}\cdot \frac{\ln
[\ln(t^{2}/\Lambda_{\rm QCD}^{2})]}{\ln(t^{2}/\Lambda_{\rm QCD}^{2})}\biggr\},
\eeq
where $\beta_{0}=(33-2N_{f})/3$ and $\beta_{1}=(306-38N_{f})/3$,
instead of the leading order (LO) elements such as LO Wilson coefficients $C_{i}(m_{W})$
and LO renormalization group evolution matrix $U(t,m)^{(0)}$, and LO
running coupling $\alpha_{s}$. For the hadronic scale $\Lambda_{\rm QCD}$, the
$\Lambda_{\rm QCD}^{(4)}=0.287$ GeV (0.326 GeV)
could be obtained by using $\Lambda_{\rm QCD}^{(5)}=0.225$ GeV
for the LO (NLO) case~\cite{Buchalla:1995vs}. For the hard scale $t$,
the lower cut-off $\mu_{0}=1.0$ GeV is chosen~\cite{Xiao:2008sw}.	
\end{itemize}

By taking various contributions from Feynman diagrams shown in Figs.~\ref{fig:fig2} and~\ref{fig:fig3}
into consideration, the decay amplitudes of $B_d^0 \to J/\psi f_0(N)$ and $B_s^0 \to J/\psi f_0(S)$
could thus be written as
\beq
\sqrt{2} A(B_{d}^0 \to J/\psi f_0(N)) &=&
F_{fe}^N f_{J/\psi}\biggl\{ V_{cb}^{\ast} V_{cd} \tilde{a}_{2}
-V_{tb}^{\ast} V_{td} \biggl( \tilde{a}_{3}+\tilde{a}_{5}
+\tilde{a}_{7}+\tilde{a}_{9}\biggr) \biggr\}
\non  &&
+M_{nfe}^N \biggl\{V_{cb}^{\ast} V_{cd} C_{2}
-V_{tb}^{\ast} V_{td}
\biggl( C_{4}-C_{6}-C_{8}+C_{10}\biggr) \biggr\}\;,
\label{eq:DecAmp-jpsifN}
\eeq
and
\beq
A(B_{s}^0 \to J/\psi f_0(S)) &=&
F_{fe}^S f_{J/\psi}\biggl\{ V_{cb}^{\ast} V_{cs} \tilde{a}_{2}
-V_{tb}^{\ast} V_{ts} \biggl( \tilde{a}_{3}+\tilde{a}_{5}
+\tilde{a}_{7}+\tilde{a}_{9}\biggr) \biggr\}
\non  &&
+M_{nfe}^S \biggl\{V_{cb}^{\ast} V_{cs} C_{2}
-V_{tb}^{\ast} V_{ts}
\biggl( C_{4}-C_{6}-C_{8}+C_{10}\biggr) \biggr\}\;,
\label{eq:DecAmp-jpsifS}
\eeq
where the four functions $(F_{fe}^N,M_{nfe}^N,F_{fe}^S,M_{nfe}^S)$ denote the factorization
formulas arising from factorizable and nonfactorizable emission diagrams in the $B_d^0 \to
J/\psi f_0(N)$ and $B_s^0 \to J/\psi f_0(S)$ modes respectively. 
The explicit forms of the factorization formulas $F_{fe}$ and $M_{nfe}$ are presented as follows,
\beq		
F_{fe} &=& -8 \pi C_{F} m^{4}_{B}\int^{1}_{0}dx_{1}dx_{3}\int^{\infty}_{0}b_{1}db_{1}b_{3}db_{3}\phi_{B}(x_{1},b_{1})\non
&\times&
\biggl\{[(r^{2}_{2}-1)(\phi_{f_0}(x_{3}) ((r^{2}_{2}-1)x_{3}-1)+r_{3}(2x_{3}-1)\phi_{f_0}^{S}(x_{3}))+r_{3}(2(r^{2}_{2}-1)x_{3}
+r^{2}_{2}+1)\phi_{f_0}^{T}(x_{3})]\non
&& \cdot h_{fe}(x_{1},x_{3},b_{1},b_{3})E_{fe}(t_{a})-[2r_{3}(r^{2}_{2}-1)\phi(x_{3})]h_{fe}(x_{3},x_{1},b_{3},b_{1})E_{fe}(t_{b})\biggl\},
\eeq
where the hard function $h_{fe}(x_i, b_i)$ and the evolution function $E_{fe}(t_i)$
could be found in Appendix~\ref{sec:app3}, and $f_0$ could be the flavor states
$N$ or $S$, respectively, corresponding to the related decays of $B_d^0$ or $B_s^0$. And,
\beq
M_{nfe}&=&-\frac{16\sqrt{6}}{3}\pi C_{F}m^{4}_{B}\int^{1}_{0}dx_{1}dx_{2}dx_{3}\int^{\infty}_{0}b_{1}db_{1}b_{2}db_{2}\phi_{B}(x_{1},b_{1}) \non
&\times&
\biggl\{[r^{2}_{2}(2x_{2}-x_{3})\phi^{L}_{J/\psi}(x_{2})+x_{3}\phi^{L}_{J/\psi}(x_{2})-2r_{2}r_{c}\phi^{t}_{J/\psi}(x_{2})]
[(r^{2}_{2}-1)\phi_{f_0}(x_{3})+2r_{3}\phi_{f_0}^{T}(x_{3})]\biggl\}
\non && \cdot
h_{nfe}(x_{1},x_{2},x_{3},b_{1},b_{2})E_{nfe}(t_{nfe}),
\eeq
where $r_c=m_c/m_{B^0}$ with $m_c$ being the charm quark mass.

According to Eq.~(\ref{eq:eigenstate}), the $B^0 \to J/\psi f_0$ decay amplitudes could then be written explicitly with
$A(B_d^0 \to J/\psi f_0(N))$ and $A(B_s^0 \to J/\psi f_0(S))$ as follows
\beq
{\cal A}(B_{d}^{0}\to J/\psi f_0)&=& A(B_{d}^{0}\to J/\psi f_0(N)) \cdot \alpha_1 \;,
\label{eq:DecAmp-psif0-d}
\\ 
{\cal A}(B_{s}^{0}\to J/\psi f_0)&=& A(B_{s}^{0}\to J/\psi f_0(S)) \cdot \alpha_2 \;,
\label{eq:DecAmp-psif0-s}
\eeq
where $\alpha_1$ and $\alpha_2$ will vary depending on the final
states $f_0(1370)$, $f_0(1500)$, and $f_0(1710)$ correspondingly.

%
%
\section{Numerical results and discussions}
\label{sect:3}
	
In this section, we will perform the numerical calculations based on the given
decay amplitudes to predict the branching fractions and their relative ratios
for the decays $B^0 \to J/\psi f_0$. Meanwhile, the phenomenological discussions on
the potential scalar glueball hunting combining with the related numerical results will be presented.
In numerical calculations, central values of the input parameters will be used implicitly unless otherwise stated.
	
First of all, several comments on the nonperturbative inputs are presented in order:
\begin{itemize}	
\item[]{(1)}
In light of the good consistency between experimental measurements and PQCD predictions
for, e.g., $B \to J/\psi V$~\cite{Liu:2013nea}, $B^0 \to J/\psi f_0(500, 980)$~\cite{Liu:2019ymi},
and so forth, the wave functions in association with the distribution amplitudes, decay constants
and mesonic masses of $B^0$ and $J/\psi$ in this work will be adopted same as those
in Refs.~\cite{Chen:2005ht,Li:2006vq,Liu:2009yno,Liu:2012ib,Liu:2013nea,Liu:2019ymi,Yao:2022zom},
but with updated $B^0$-meson lifetimes, i.e.,
$\tau_{B_d^0}= 1.519 \ \rm ps$ and $\tau_{B_s^0}= 1.516\  \rm ps$~\cite{ParticleDataGroup:2020ssz}.

\item[]{(2)}
For the wave functions and distribution amplitudes of the scalar flavor states $N$ and $S$,
we also take the same forms as those in the studies of $B^0 \to J/\psi f_0(500, 980)$
\cite{Liu:2019ymi}, but replacing the masses, decay constants and Gegenbauer moments
of $f_0$~\cite{Cheng:2005nb, Li:2008tk} considered in this work correspondingly. 
Notice that, there are no any available decay constants and Gegenbauer moments for the flavor states $N$ and $S$ in the literature currently. To our best knowledge, the available decay constants and Gegenbauer moments in the twist-2 and tiwst-3 distribution amplitudes for the light scalars $a_0(1450), K_0^*(1430)$, and $f_0(1500)$ have been calculated in the picture of $q\bar q$-meson with QCD sum rules~\cite{Cheng:2005nb,Lu:2006fr,Han:2013zg}. Furthermore, in the past two decades, the scalar $f_0(1500)$ was always assumed as the almost pure $s\bar s$ state in the relevant investigations with factorization frameworks based on QCD. The very recent study could be found in~\cite{Han:2023pgf}. Therefore, in this work, we will take the decay constant and the Gegenbauer moments of $S$ as those of $f_0(1500)$.
And for the less
understood $f_0(1370)$, as aforementioned in Sect.~\ref{sect:2}, it is predominated by $N$ according to the reached consensus. Then, like $f_\omega \sim f_\rho$ in the vector sector~\cite{Ali:2007ff} and $f_{\eta_q} \sim f_\pi$ in the pseudoscalar sector~\cite{Charng:2006zj}, we adopt the decay constant of $N$ as that of $a_0(1450)$ in the limit of isopsin symmetry. Furthermore, based on the light-cone distribution amplitudes in Eqs.~(\ref{eq:lcda-fq-t2})-(\ref{eq:lcda-t-fq-t3g}), the decay constants could be factored out from the factorization formulas, which would be cancelled in the ratios between the branching ratios of the related channels. Moreover, as for the Gegenbauer moments in the distribution amplitudes of $N$, the relation $B_{1,3}^S \simeq 0.8 B_{1,3}^N$ is adopted~\footnote{\RyB{It is worth emphasizing that this relation has been used in the investigations of $B^0 \to J/\psi f_0(500, 980)$~\cite{Liu:2019ymi}. And the PQCD results are in global consistency with the current measurements.}}, according to the values for vacuum condensates $<s\bar s> = 0.8 <u\bar u>$~\cite{Cheng:2005ye,Lu:2006fr} (For details, see the appendices~A and B in~\cite{Cheng:2005ye}). Therefore, the values of the decay constants and the
Gegenbauer moments for the flavor states $N$ and $S$ in QCD sum rules at the renormalization scale $\mu =1$~GeV are collected as follows,
\begin{itemize}
\item[] {(a)}
For the scalar decay constants (in units of GeV) of
$N$ and $S$, \Gre{we have}
\beq
\bar f_{N} &=& \biggl\{\begin{array}{cc} -0.280\pm 0.030  \qquad ({\rm S1})  \\ \hspace{3.2mm}
0.460\pm 0.050  \qquad ({\rm S2}) \end{array} \;, \label{eq:DecConst0} \\
\bar f_{S} &=& \biggl\{\begin{array}{cc} -0.255\pm 0.030 \qquad ({\rm S1})\\
\hspace{3.2mm} 0.490\pm 0.050 \qquad  ({\rm S2}) \end{array} \;. \label{eq:DecConst}
\eeq
Notice that, because of the charge conjugation invariance or the conservation of the vector current,
the neutral scalars cannot be produced via vector current, which consequently results in the zero
vector decay constants, i.e., $f_{N}= f_{S}= 0$~\cite{Cheng:2005nb}.
		
\item[] {(b)}
For the Gegenbauer moments $B_1$ and $B_3$ in leading-twist distribution amplitudes, we
\beq
B_1^N &=& \biggl\{\begin{array}{cc}\hspace{3.2mm} 1.00\pm 0.50 \qquad   \ ({\rm S1})  \\ -0.60\pm 0.14 \qquad   \ ({\rm S2}) \end{array} \;, \non
B_1^S &=& \biggl\{\begin{array}{cc} \hspace{3.2mm}
0.80\pm 0.40 \qquad   \ ({\rm S1})\\
-0.48\pm 0.11 \qquad   \ ({\rm S2})
\end{array}\;, \label{eq:GegenMomts-B1}
\\
B_3^N &=& \biggl\{\begin{array}{cc}-1.65\pm 0.18  \qquad   \ ({\rm S1})   \\ -0.46\pm 0.25  \qquad   \ ({\rm S2}) \end{array} \;,
\non
B_3^S &=& \biggl\{\begin{array}{cc} -1.32\pm 0.14 \qquad   \ ({\rm S1})\\ -0.37\pm 0.20\qquad   \ ({\rm S2})
\end{array}\;.
\label{eq:GegenMomts-B3}
\eeq
\end{itemize}
Moreover, for masses of the scalar quarkonia and the physical states,
we adopt the values (in units of GeV) as:
$m_{f_0(1370)}=1.350$, $m_{f_0(1500)}=1.506$, $m_{f_0(1710)}=1.704$~\cite{ParticleDataGroup:2020ssz},
$m_{N} = 1.304$ and $m_{S} = 1.682$ in $M_{\rm I}$~\cite{Close:2005vf},
and $m_{N} = 1.474$ and $m_{S} = 1.496$ in $M_{\rm II}$~\cite{Cheng:2015iaa}, respectively.
The necessary remark for $m_{f_0(1370)}$ is that
we approximately take the central value $1350$~MeV based on the data,
namely, 1200-1500~MeV provided by PDG, though the very existence of $f_0(1370)$ has long been
questionable (see Refs.~\cite{Bugg:2007ja,Klempt:2007cp,Ochs:2013gi} for more detailed discussions)
\footnote{\RyB{ A latest work~\cite{Pelaez:2022qby} confirmed the existence of resonance
$f_0(1370)$ in the dispersive analyses
of meson-meson scattering data and found its pole at $(1245 \pm 40) -  i (300^{+30}_{-70})$~MeV
in the $\pi\pi \to \pi\pi$ amplitude. Meanwhile, a pole at $(1380^{+70}_{-60}) -i (220^{+80}_{-70})$~MeV
also appeared in the $\pi\pi \to K\bar K$ data analysis with partial-wave dispersion relations.
}}.

\item[]{(3)}
For the CKM matrix elements, we adopt the Wolfenstein parameterization up
to $\mathcal{O}(\lambda^{5})$ \cite{Wolfenstein:1983yz}
with the four updated parameters~\cite{ParticleDataGroup:2020ssz},
\beq
A=0.790, \quad \lambda=0.22650,
\quad \bar\rho=0.141_{-0.017}^{+0.016}, \quad \bar\eta=0.357_{-0.011}^{+0.011}.
\eeq
\end{itemize}
	

With the decay amplitudes ${\cal A}(B^0 \to J/\psi f_0)$ given in Eqs.~(\ref{eq:DecAmp-psif0-d})
and~(\ref{eq:DecAmp-psif0-s}), the branching fraction for the decay $B^0 \to J/\psi f_0$
is given as follows,
\beq
{\cal B}(B^0 \to J/\psi f_0)&\equiv&
\tau_{B^0}\cdot \Gamma(B^0 \to J/\psi f_0)
\non
&=& \tau_{B^0}\cdot\frac{G_{F}^{2} m^{3}_{B^0}}
{32 \pi}\cdot \Phi(r_{J/\psi}, r_{f_0})
\cdot |{\cal A}(B^0\to J/\psi f_0) |^2\;,
\label{eq:br-def}
\eeq
where $\tau_{B^0}$ is the lifetime of $B^0$-meson and $\Phi(r_{J/\psi}, r_{f_0})
$ stands for the phase space factor of $B^0 \to J/\psi f_0$
with $r_{J/\psi} = m_{J/\psi}/m_{B^0}$ and $r_{f_0} = m_{f_0}/m_{B^0}$,
$\Phi(x, y) \equiv \sqrt{ [1-(x+y)^2] \cdot[1-(x-y)^2] }$~\cite{Fleischer:2011au}.

Now we calculate the $B^0 \to J/\psi f_0$ branching fractions
in PQCD approach at NLO level. The numerical results
within two different scenarios
in both models I and II are presented in Table~\ref{tab:BR-MI&II}.
Furthermore, we sequentially list the dominant
errors arising from theoretical uncertainties
of the shape parameter $\omega_{B_d^0} = 0.40 \pm 0.04$~GeV
or $\omega_{B_s^0} = 0.50 \pm 0.05$~GeV in the $B^0$-meson
distribution amplitude, of the $J/\psi$-meson decay constant $f_{J/\psi} = 0.405 \pm 0.014$~GeV,
of the Gegenbauer moments $B_i^{N, S} (i=1,3)$ and the scalar decay constants $\bar f_{N, S}$ (see Eqs.~(\ref{eq:DecConst})-(\ref{eq:GegenMomts-B3}))
in the light-cone distribution amplitudes of scalar quarkonia $N$ and $S$,
and of the possibly higher order contributions by simply varying
the running hard scale $t_{max}$, i.e., from $0.8t$ to $1.2t$, in the hard kernel.
\begin{table}[htb]
\caption{
The branching fractions for $B^0 \to J/\psi f_0$ with PQCD approach
in both models I and II.
The upper (lower) entry corresponds to $f_0$ in scenario 1 (2)
at every line.}
\label{tab:BR-MI&II}
\begin{center}\vspace{-0.5cm}{
\begin{tabular}[t]{c|c|c}
\hline  \hline
Decay modes   &  model I & model II \\
\hline \hline
$B_d^0 \to J/\psi f_0(1370)$
&$\begin{array}{cc}
(8.54^{+3.44+0.50+6.67+0.65+1.79}_{-2.48-0.49-1.89-1.27-1.62})\times 10^{-6}
\\
(3.65^{+1.17+0.47+1.62+0.17+0.77}_{-0.84-0.42-1.41-0.27-0.69})\times 10^{-5}
\end{array}$
&$\begin{array}{cc}
(1.04^{+0.43+0.06+0.74+0.08+0.22}_{-0.31-0.06-0.23-0.15-0.20})\times 10^{-5}
\\
(4.38^{+1.44+0.57+1.90+0.22+0.92}_{-1.03-0.52-1.66-0.36-0.83})\times 10^{-5}
\end{array}$
\\
\hline
$B_d^0 \to J/\psi f_0(1500)$
&$\begin{array}{cc}
(3.55^{+1.43+0.21+2.78+0.27+0.75}_{-1.03-0.20-0.79-0.53-0.67})\times 10^{-6}
\\
(1.52^{+0.48+0.19+0.67+0.07+0.32}_{-0.35-0.18-0.59-0.12-0.29})\times 10^{-5}
\end{array}$
&$\begin{array}{cc}
(2.10^{+0.87+0.12+1.49+0.15+0.43}_{-0.62-0.13-0.47-0.31-0.40})\times 10^{-6}
\\
(8.84^{+2.89+1.15+3.82+0.44+1.83}_{-2.09-1.05-3.36-0.73-1.67})\times 10^{-6}
\end{array}$
\\
\hline
$B_d^0 \to J/\psi f_0(1710)$
&$\begin{array}{cc}
(2.93^{+1.18+0.18+2.29+0.23+0.62}_{-0.85-0.16-0.64-0.43-0.56})\times 10^{-7}
\\
(1.26^{+0.39+0.16+0.55+0.05+0.26}_{-0.30-0.15-0.49-0.10-0.24})\times 10^{-6}
\end{array}$
&$\begin{array}{cc}
(6.73^{+2.80+0.41+4.80+0.51+1.43}_{-1.98-0.39-1.49-1.00-1.29})\times 10^{-7}
\\
(2.84^{+0.93+0.37+1.23+0.14+0.60}_{-0.67-0.34-1.08-0.23-0.55})\times 10^{-6}
\end{array}$
\\
\hline \hline
$B_s^0 \to J/\psi f_0(1370)$
&$\begin{array}{cc}
(5.07^{+2.97+0.29+3.62+0.40+1.03}_{-1.90-0.27-1.23-0.77-0.94})\times 10^{-6}
\\
(3.38^{+1.35+0.47+1.41+0.31+0.69}_{-0.92-0.42-1.25-0.36-0.63})\times 10^{-5}
\end{array}$
&$\begin{array}{cc}
(1.90^{+1.09+0.11+1.53+0.16+0.40}_{-0.69-0.09-0.45-0.29-0.36})\times 10^{-5}
\\
(1.30^{+0.51+0.18+0.30+0.11+0.27}_{-0.35-0.10-0.49-0.13-0.25})\times 10^{-4}
\end{array}$
\\
\hline
$B_s^0 \to J/\psi f_0(1500)$
&$\begin{array}{cc}
(9.25^{+5.42+0.53+6.63+0.73+1.94}_{-3.46-0.50-2.25-1.39-1.76})\times 10^{-6}
\\
(6.17^{+2.47+0.85+2.57+0.56+1.29}_{-1.67-0.77-2.28-0.66-1.17})\times 10^{-5}
\end{array}$
&$\begin{array}{cc}
(1.22^{+0.69+0.06+0.97+0.10+0.25}_{-0.45-0.06-0.29-0.19-0.23})\times 10^{-4}
\\
(8.32^{+3.23+1.12+3.55+0.69+1.74}_{-2.22-0.65-3.13-0.83-1.59})\times 10^{-4}
\end{array}$
\\
\hline
$B_s^0 \to J/\psi f_0(1710)$
&$\begin{array}{cc}
(1.36^{+0.80+0.08+0.99+0.11+0.29}_{-0.51-0.07-0.33-0.20-0.25})\times 10^{-4}
\\
(9.10^{+3.64+1.25+3.80+0.83+1.90}_{-2.46-1.14-3.36-0.98-1.73})\times 10^{-4}
\end{array}$
&$\begin{array}{cc}
(1.07^{+0.62+0.06+0.86+0.09+0.23}_{-0.39-0.05-0.25-0.16-0.20})\times 10^{-5}
\\
(7.32^{+2.86+0.99+3.14+0.62+1.57}_{-1.95-0.57-2.74-0.73-1.41})\times 10^{-5}
\end{array}$
\\
\hline \hline
\end{tabular}}
\end{center}
\end{table}
The last errors in the second and third columns of Table~\ref{tab:BR-MI&II}
stem from $10\%$ variations of the related coefficients for scalar quarkonia $N$ and $S$ in the mixing matrices.
Due to their very smallness, the errors induced by the CKM matrix elements
are negligible and not shown in the Table. It is obvious that the largest uncertainty arises from the least
constrained hadronic parameters, i.e., the Gegenbauer moments $B_i^{N, S}(i=1,3)$ and the
scalar decay constants $\bar f_{N, S}$, which are nonperturbative.
Nevertheless, such large branching fractions of the decays $B^0 \to J/\psi f_0$ within
still large uncertainties generally around $10^{-6} \sim 10^{-3}$
would provide good probes to these parameters,
because of the highly small
penguin contributions. These numerical results are expected to be tested at LHCb and Belle-II experiments in the near future,
which might provide useful constraints on the nonperturbative inputs and further lead to
more reliable predictions.
In principle, the future experimental confirmations on these large predictions
could be helpful to determine the coefficients of the scalar quarkonia, namely,
$\alpha_1$ for $N$ and $\alpha_2$ for $S$ in three considered scalars $f_0$.
	
\begin{itemize}	
\item[(1)]{The decays $B^0 \to J/\psi f_0(1370)$ }

\hspace{0.5mm}
As seen from the PQCD branching fractions in Table~\ref{tab:BR-MI&II},
the consensus about dominated quarkonium $N$ of $f_0(1370)$ in both models I and II leads to
the consistent ${\cal B}(B_d^0 \to J/\psi f_0(1370))$ within uncertainties in each scenario, namely,
\beq
{\cal B}(B_d^0 \to J/\psi f_0(1370))_{\rm S1} &=&
\biggl\{\begin{array}{ll}
(0.85^{+0.78}_{-0.38}) \times 10^{-5} \qquad  \ (M_{\rm I}) \\
(1.04^{+0.89}_{-0.47}) \times 10^{-5} \qquad  \ (M_{\rm II})
\end{array} \;,
\label{eq:psif13-d-s1}\\
{\cal B}(B_d^0 \to J/\psi f_0(1370))_{\rm S2} &=&
\biggl\{\begin{array}{ll}
(3.65^{+2.20}_{-1.85}) \times 10^{-5} \qquad  \ (M_{\rm I}) \\
(4.38^{+2.63}_{-4.92}) \times 10^{-5} \qquad  \ (M_{\rm II})
\end{array} \;,
\label{eq:psif13-d-s2}
\eeq
which could be inferred from the close values $|\alpha_{1}^{M_{\rm I}}| = 0.810 \pm 0.081$ and
$|\alpha_{1}^{M_{\rm II}}| = 0.862 \pm 0.086$ presented in
Eqs.~(\ref{eq:MI}) and~(\ref{eq:MII}).
These large results for the branching fractions of the decay $B_d^0 \to
J/\psi f_0(1370)$ are around $10^{-6} \sim 10^{-5}$ and could be accessible
at the LHCb and Belle-II experiments.
Meanwhile, it is worth mentioning that, though the $f_0(1370)$ is predominated by the
scalar quarkonium $N$, the branching fractions of the decay $B_s^0 \to J/\psi f_0(1370)$
are still large,
\beq
{\cal B}(B_s^0 \to J/\psi f_0(1370))_{\rm S1} &=&
\biggl\{\begin{array}{ll}
(0.51^{+0.48}_{-0.26}) \times 10^{-5} \qquad  \ (M_{\rm I}) \\
(1.90^{+1.93}_{-0.95}) \times 10^{-5} \qquad  \ (M_{\rm II})
\end{array} \;,
\label{eq:psif13-s-s1}\\
{\cal B}(B_s^0 \to J/\psi f_0(1370))_{\rm S2} &=&
\biggl\{\begin{array}{ll}
(3.38^{+2.14}_{-1.76}) \times 10^{-5} \qquad  \ (M_{\rm I}) \\
(1.30^{+0.68}_{-0.67}) \times 10^{-4} \qquad  \ (M_{\rm II})
\end{array} \;,
\label{eq:psif13-s-s2}
\eeq
which are capable of measuring
at the on-going LHCb and Belle-II experiments.

\hspace{0.5mm}
It is emphasized that, as presented in Eq.~(\ref{eq:psif13s-pp-ex-lhcb}),
this mode has been reported through
$f_0(1370) \to \pi^+ \pi^-$ while with surprisingly large uncertainties in
the LHCb measurements, besides the evidence given by the Belle collaboration.
As stated in~\cite{Cheng:2015iaa}, the narrow-width approximation (NWA)
works provided that the resonance is not too broad.
But, experimentally, the nature of $f_0(1370)$ is
unknown till now. Thus, it looks unfeasible to
make an efficient comparison
straightforwardly for the branching fractions between the PQCD predictions
and the LHCb measurements under NWA.
Therefore, for future tests at relevant experiments, by taking the decays $B_d^0 \to J/\psi \rho^0$
and $B_s^0 \to J/\psi \phi$ as normalization, we can define the ratios
between the branching fractions \RyB{of} $B^0 \to J/\psi f_0$ and $B^0 \to J/\psi \rho^0/ \phi$.
The updated values for ${\cal B}(B^0_d \to J/\psi \rho^0)$ and ${\cal B}(B^0_s \to J/\psi \phi)$
in PQCD are available as $(2.98^{+0.81}_{-0.69})\times 10^{-5}$
and $(1.07^{+0.33}_{-0.29})\times 10^{-3}$~\cite{Yao:2022zom} correspondingly.
The values for the two ratios, namely, $R_d^{\rm Theo}[f_{03}/\rho]$
and $R_s^{\rm Theo}[f_{03}/\phi]$, with theoretical errors
are then collected as,
\beq
R_d^{\rm Theo}[f_{03}/\rho]_{\rm S1}
&\equiv&
\frac{{\cal B}(B_d^0 \to J/\psi f_0(1370))_{\rm S1}}
{{\cal B}(B_d^0 \to J/\psi \rho^0)}
=
\biggl\{\begin{array}{ll}
0.287^{+0.201}_{-0.072}  \qquad  \ (M_{\rm I}) \\
0.349^{+0.222}_{-0.089}  \qquad  \ (M_{\rm II})
\end{array} \;,
\label{eq:f13rho-d-s1}
\\
R_d^{\rm Theo}[f_{03}/\rho]_{\rm S2}
&\equiv&
\frac{{\cal B}(B_d^0 \to J/\psi f_0(1370))_{\rm S2}}
{{\cal B}(B_d^0 \to J/\psi \rho^0)}
=
\biggl\{\begin{array}{ll}
1.225^{+0.461}_{-0.434} \qquad  \ (M_{\rm I}) \\
1.470^{+0.542}_{-0.512} \qquad  \ (M_{\rm II})
\end{array} \;,
\label{eq:f13rho-d-s2}
\eeq
and
\beq
R_s^{\rm Theo}[f_{03}/\phi]_{\rm S1}
&\equiv&
\frac{{\cal B}(B_s^0 \to J/\psi f_0(1370))_{\rm S1}}
{{\cal B}(B_s^0 \to J/\psi \phi)}
=
\biggl\{\begin{array}{ll}
0.005^{+0.003}_{-0.001}  \qquad  \ (M_{\rm I}) \\
0.018^{+0.013}_{-0.005}  \qquad  \ (M_{\rm II})
\end{array} \;,
\label{eq:f13phi-s-s1}
\\
R_s^{\rm Theo}[f_{03}/\phi]_{\rm S2}
&\equiv&
\frac{{\cal B}(B_s^0 \to J/\psi f_0(1370))_{\rm S2}}
{{\cal B}(B_s^0 \to J/\psi \phi)}
=
\biggl\{\begin{array}{ll}
0.032^{+0.011}_{-0.010} \qquad  \ (M_{\rm I}) \\
0.121^{+0.045}_{-0.041} \qquad  \ (M_{\rm II})
\end{array} \;.
\label{eq:f13phi-s-s2}
\eeq
The large ratios such as $R_d^{\rm Theo}[f_{03}/\rho]_{\rm S2, M_{I}} \in [0.791, 1.686]$
and $R_d^{\rm Theo}[f_{03}/\rho]_{\rm S2, M_{II}} \in [0.958, 2.012]$, and
$R_s^{\rm Theo}[f_{03}/\phi]_{\rm S2, M_{II}} \in [0.080, 0.166]$
need more experimental tests for further understanding the nature
of $f_0(1370)$.

\item[(2)]{The decays $B^0 \to J/\psi f_0(1500)$}

\hspace{0.5mm}	
Based on the results presented in Table~\ref{tab:BR-MI&II},
the PQCD predictions for ${\cal B}(B^0 \to J/\psi f_0(1500))$ are collected as follows,
\beq
{\cal B}(B_d^0 \to J/\psi f_0(1500))_{\rm S1} &=&
\biggl\{\begin{array}{ll}
(3.55^{+3.23}_{-1.57}) \times 10^{-6} \qquad  \ (M_{\rm I}) \\
(2.10^{+1.79}_{-0.94}) \times 10^{-6} \qquad  \ (M_{\rm II})
\end{array} \;,
\label{eq:psif15-d-s1}\\
{\cal B}(B_d^0 \to J/\psi f_0(1500))_{\rm S2} &=&
\biggl\{\begin{array}{ll}
(1.52^{+0.91}_{-0.78}) \times 10^{-5} \qquad  \ (M_{\rm I}) \\
(8.84^{+5.27}_{-4.48}) \times 10^{-6} \qquad  \ (M_{\rm II})
\end{array} \;,
\label{eq:psif15-d-s2}
\eeq
and
\beq
{\cal B}(B_s^0 \to J/\psi f_0(1500))_{\rm S1} &=&
\biggl\{\begin{array}{ll}
(0.93^{+0.88}_{-0.47}) \times 10^{-5} \qquad  \ (M_{\rm I}) \\
(1.22^{+1.22}_{-0.62}) \times 10^{-4} \qquad  \ (M_{\rm II})
\end{array} \;,
\label{eq:psif15-s-s1}\\
{\cal B}(B_s^0 \to J/\psi f_0(1500))_{\rm S2} &=&
\biggl\{\begin{array}{ll}
(6.17^{+3.93}_{-3.22}) \times 10^{-5} \qquad  \ (M_{\rm I}) \\
(8.32^{+5.27}_{-4.29}) \times 10^{-4} \qquad  \ (M_{\rm II})
\end{array} \;,
\label{eq:psif15-s-s2}
\eeq
where the uncertainties from various sources have been added in quadrature.
In principle, these large branching fractions around $10^{-6} \sim 10^{-4}$
are expected to be tested soon at the LHCb and Belle-II
experiments. It is mentioned that, due to the comparable $N$-component of $f_0(1500)$ in
models I and II, the $B_d^0 \to J/\psi f_0(1500)$ decay rates are generally consistent with each other in each scenario.
However, for the $B_s^0$ decay mode, the relation ${\cal B}(B_s^0 \to J/\psi f_0(1500))_{M_{\rm II}} \sim 10 \times
{\cal B}(B_s^0 \to J/\psi f_0(1500))_{M_{\rm I}}$ could be seen in both S1 and S2. It is indeed attributed to the
fact that the $f_0(1500)$ state is assumed as the prominent glueball nature in model I while as the flavor octet structure
in model II.

\hspace{0.5mm}
The ratios $R_d^{\rm Theo}[f_{05}/\rho]$ and $R_s^{\rm Theo}[f_{05}/\phi]$ are then presented
similarly as,
\beq
R_d^{\rm Theo}[f_{05}/\rho]_{\rm S1}
&\equiv&
\frac{{\cal B}(B_d^0 \to J/\psi f_0(1500))_{\rm S1}}
{{\cal B}(B_d^0 \to J/\psi \rho^0)}
=
\biggl\{\begin{array}{ll}
0.119^{+0.084}_{-0.030}  \qquad  \ (M_{\rm I}) \\
0.070^{+0.044}_{-0.018}  \qquad  \ (M_{\rm II})
\end{array} \;,
\label{eq:f15rho-d-s1}
\\
R_d^{\rm Theo}[f_{05}/\rho]_{\rm S2}
&\equiv&
\frac{{\cal B}(B_d^0 \to J/\psi f_0(1500))_{\rm S2}}
{{\cal B}(B_d^0 \to J/\psi \rho^0)}
=
\biggl\{\begin{array}{ll}
0.510^{+0.190}_{-0.181} \qquad  \ (M_{\rm I}) \\
0.297^{+0.109}_{-0.103} \qquad  \ (M_{\rm II})
\end{array} \;,
\label{eq:f15rho-d-s2}
\eeq
and
\beq
R_s^{\rm Theo}[f_{05}/\phi]_{\rm S1}
&\equiv&
\frac{{\cal B}(B_s^0 \to J/\psi f_0(1500))_{\rm S1}}
{{\cal B}(B_s^0 \to J/\psi \phi)}
=
\biggl\{\begin{array}{ll}
0.009^{+0.006}_{-0.003}  \qquad  \ (M_{\rm I}) \\
0.114^{+0.083}_{-0.033}  \qquad  \ (M_{\rm II})
\end{array} \;,
\label{eq:f15phi-s-s1}
\\
R_s^{\rm Theo}[f_{05}/\phi]_{\rm S2}
&\equiv&
\frac{{\cal B}(B_s^0 \to J/\psi f_0(1500))_{\rm S2}}
{{\cal B}(B_s^0 \to J/\psi \phi)}
=
\biggl\{\begin{array}{ll}
0.058^{+0.021}_{-0.019} \qquad  \ (M_{\rm I}) \\
0.778^{+0.282}_{-0.255} \qquad  \ (M_{\rm II})
\end{array} \;,
\label{eq:f15phi-s-s2}
\eeq
which could be helpful to study the inner structure of $f_0(1500)$ and to explore the fraction of glueball content.

\hspace{0.5mm}
Actually, as mentioned in Sect.~\ref{sect:1}, the LHCb Collaboration reported the experimental result
of ${\cal B}(B_s^0 \to J/\psi f_0(1500), f_0 \to \pi^+ \pi^-)$ as $2.04^{+0.32}_{-0.24}\times 10^{-5}$.
Then the  ${\cal B}(B_s^0 \to J/\psi f_0(1500))_{\rm Exp}$ could be easily derived with the data
${\cal B}(f_0(1500) \to \pi\pi) = (34.5\pm 2.2)\%$~\cite{ParticleDataGroup:2020ssz}
under NWA~\cite{Cheng:2015iaa}. With ${\cal B}(f_0(1500) \to \pi^+\pi^-) = 0.230 \pm 0.015$
based on the isospin symmetry, we have
\beq
{\cal B}(B_s^0 \to J/\psi f_0(1500))_{\rm Exp} &=& (8.87^{+0.76}_{-0.50}) \times 10^{-5}\;.
\label{eq:psif15s-ex}
\eeq
Experimentally, the $B_s^0 \to J/\psi \phi$ branching fraction~\cite{ParticleDataGroup:2020ssz} is read as,
\beq
{\cal B}(B_s^0 \to J/\psi \phi)_{\rm Exp} &=& (1.08^{+0.08}_{-0.08}) \times 10^{-3}\;.
\label{eq:psiphi-ex}
\eeq
Therefore, the ratio between the experimental branching fractions
of $B_s^0 \to J/\psi f_0(1500)$ and $B_s^0 \to J/\psi \phi$ could be
obtained as,
\beq
R_{s}^{\rm Exp}[f_{05}/\phi] &\equiv&
\frac{{\cal B}(B_s^0 \to J/\psi f_0(1500))_{\rm Exp}}{{\cal B}(B_s^0 \to J/\psi \phi)_{\rm Exp}}
= 0.082^{+0.002}_{-0.000} \;.
\label{eq:r-f15-phi-ex}
\eeq
By combining the branching fractions and the relative ratios, the results
in the scheme of S1 with $M_{\rm II}$ seem closer to the values
in Eqs.~(\ref{eq:psif15s-ex}) and~(\ref{eq:r-f15-phi-ex}) derived from
data than those in the scheme of S2 with $M_{\rm I}$. However,
due to the still large uncertainties, the results in both of the above mentioned two schemes,
i.e., S1 with $M_{\rm II}$ and S2 with $M_{\rm I}$, are consistent with the current data in 2$\sigma$ deviations.
Certainly, it looks evidently that the values of
${\cal B}(B_s^0 \to J/\psi f_0(1500))$ in both schemes of S1 with $M_{\rm I}$ and S2 with $M_{\rm II}$
are less favored by that in Eq.~(\ref{eq:psif15s-ex}). The crosschecks from the Belle-II
experiments for the decay $B_s^0 \to J/\psi f_0(1500) (\to \pi^+ \pi^-)$ are thus urgently needed.

\hspace{0.5mm}
For comparing with the future measurements, we read the branching fractions
for the decays $B^0 \to J/\psi f_0(1500)( \to \pi^+ \pi^-)$
by using NWA~\cite{Cheng:2015iaa}. These theoretical predictions are collected in Table~\ref{tab:BR4psif15-NWA}.
Notice that the values for ${\cal B}(B_s^0 \to J/\psi f_0(1500) (\to \pi^+ \pi^-))$
in both schemes of S1 with $M_{\rm II}$ and S2 with $M_{\rm I}$ are well consistent with the
current LHCb measurement within theoretical errors.
The branching fractions in the order of $10^{-6}$ for the
mode $B_d^0 \to J/\psi f_0(1500) (\to \pi^+ \pi^-)$ are accessible in the on-going experiments
at Belle-II and LHCb and the predictions await the near future examinations.
\begin{table}[htb]
\caption{ The branching fractions for the decays
$B^0 \to J/\psi f_0(1500)(\to \pi^+\pi^-/ K^+ K^-)$ under NWA in the PQCD approach.
The upper (lower) entry corresponds to $f_0(1500)$ in scenario 1 (2)
at every line. }
\label{tab:BR4psif15-NWA}
\begin{center}\vspace{-0.5cm}{
\begin{tabular}[t]{c|c|c}
\hline  \hline
Decay modes   &  model I  & model II \\
\hline \hline
$B_d^0 \to J/\psi f_0(1500)(\to \pi^+ \pi^-)$
&$\begin{array}{cc}
(8.17^{+7.45}_{-3.64})\times 10^{-7}
\\
(3.50^{+2.11}_{-1.80})\times 10^{-6}
\end{array}$
&$\begin{array}{cc}
(4.83^{+4.12}_{-2.18})\times 10^{-7}
\\
(2.03^{+1.22}_{-1.04})\times 10^{-6}
\end{array}$
\\
\hline
$B_s^0 \to J/\psi f_0(1500)(\to \pi^+ \pi^-)$
&$\begin{array}{cc}
(2.13^{+2.04}_{-1.10})\times 10^{-6}
\\
(1.42^{+0.91}_{-0.75})\times 10^{-5}
\end{array}$
&$\begin{array}{cc}
(2.81^{+2.81}_{-1.43})\times 10^{-5}
\\
(1.91^{+1.22}_{-0.99})\times 10^{-4}
\end{array}$
\\
\hline \hline
$B_d^0 \to J/\psi f_0(1500)(\to K^+ K^-)$
&$\begin{array}{cc}
(1.51^{+1.39}_{-0.69})\times 10^{-7}
\\
(6.46^{+3.94}_{-3.39})\times 10^{-7}
\end{array}$
&$\begin{array}{cc}
(8.93^{+7.67}_{-4.13})\times 10^{-8}
\\
(3.76^{+2.29}_{-1.96})\times 10^{-7}
\end{array}$
\\
\hline
$B_s^0 \to J/\psi f_0(1500)(\to K^+ K^-)$
&$\begin{array}{cc}
(3.93^{+3.78}_{-2.06})\times 10^{-7}
\\
(2.62^{+1.70}_{-1.41})\times 10^{-6}
\end{array}$
&$\begin{array}{cc}
(5.19^{+5.22}_{-2.69})\times 10^{-6}
\\
(3.54^{+2.28}_{-1.87})\times 10^{-5}
\end{array}$
\\
\hline \hline
\end{tabular}}
\end{center}
\end{table}

\hspace{0.5mm}
Analogously, as byproducts, the results for ${\cal B}(B^0 \to
J/\psi f_0(1500) (\to K^+ K^-))$ could also be obtained with the data ${\cal B}(f_0(1500)
\to K^+ K^-)=(4.25 \pm 0.50)\%$~\cite{ParticleDataGroup:2020ssz} on the basis of isospin
symmetry. Using ${\cal B}(B_s^0 \to J/\psi f_0(1500))_{\rm Exp}$ and
${\cal B}(f_0(1500) \to K^+ K^-)_{\rm Exp}$, the future measurement about
${\cal B}(B_s^0 \to J/\psi f_0(1500) (\to K^+ K^-))$ might
be as approximately described under NWA,
\beq
{\cal B}(B_s^0 \to J/\psi f_0(1500) (\to K^+ K^-) )_{\rm Exp}
&\simeq&
{\cal B}(B_s^0 \to J/\psi f_0(1500))_{\rm Exp} \cdot
{\cal B}(f_0(1500) \to K^+ K^-)_{\rm Exp}
\non
&=&
(3.77^{+0.80}_{-0.63}) \times 10^{-6}\;.
\label{eq:psif15s-kk-ex}
\eeq
Meanwhile, the PQCD predictions around $10^{-6} \sim 10^{-5}$ within uncertainties for
${\cal B}(B_s^0 \to J/\psi f_0(1500)\\ (\to K^+ K^-))$ can be seen in Table~\ref{tab:BR4psif15-NWA}.
Furthermore, we can find that both of the numerical results in S1 with $M_{\rm II}$
and in S2 with $M_{\rm I}$ are well consistent with those as shown in Eq.~(\ref{eq:psif15s-kk-ex})
within theoretical errors. They could be tested in the experiments at LHC and KEK.
Notice that, here, $f_0(1500)$ is assumed as primary scalar
glueball in model I while as predominant scalar quarkonium $S$ in model II. If the future measurements confirm
the result in Eq.~(\ref{eq:psif15s-kk-ex}) and the consistency between it and the
PQCD predictions, it will imply distinct couplings $g_{f_0 K K}$ of scalar glueball in
different models, which may be the key point to differentiate the possible scalar glueball.

\begin{table}[htb]
\caption{The PQCD predicted ratios between the branching fractions of $B^0 \to J/\psi f_0(1500) (\to \pi^+ \pi^-/ K^+K^-)$
and $B^0 \to J/\psi \rho^0/\phi (\to \pi^+ \pi^-/ K^+K^-)$ in both models I and II.
The upper (lower) entry corresponds to $f_0(1500)$ in scenario 1 (2) at every line. }
\label{tab:Rs-BR-15}
\begin{center}\vspace{-0.5cm}{
\begin{tabular}[t]{c|c|c}
\hline  \hline
Ratios   &  model I  & model II \\
\hline \hline
$R_{d, \pi\pi}^{\rm Theo}[f_{05}/\rho] \equiv
\frac{{\cal B}(B_d^0 \to J/\psi f_0(1500) (\to \pi^+ \pi^-))}
{{\cal B}(B_{d}^0 \to J/\psi \rho^0 (\to \pi^{+} \pi^{-}))}$
&$\begin{array}{cc}
0.027^{+0.019}_{-0.007} \\
0.117^{+0.044}_{-0.042}
\end{array}$
&$\begin{array}{cc}
0.016^{+0.010}_{-0.004} \\
0.068^{+0.025}_{-0.024}
\end{array}$
\\
\hline
$R_{d, K \pi}^{\rm Theo}[f_{05}/\rho] \equiv
\frac{{\cal B}(B_d^0 \to J/\psi f_0(1500) (\to K^+ K^-))}
{{\cal B}(B_{d}^0 \to J/\psi \rho^0 (\to \pi^{+} \pi^{-}))}$
&$\begin{array}{cc}
0.005^{+0.004}_{-0.001}  \\
0.022^{+0.009}_{-0.008}
\end{array}$
&$\begin{array}{cc}
0.003^{+0.002}_{-0.001}  \\
0.013^{+0.005}_{-0.004}
\end{array}$
\\
\hline
$R_{s, \pi K}^{\rm Theo}[f_{05}/\phi] \equiv
\frac{{\cal B}(B_s^0 \to J/\psi f_0(1500) (\to \pi^+ \pi^-))}
{{\cal B}(B_{s}^0 \to J/\psi \phi (\to K^{+} K^{-})) }$
&$\begin{array}{cc}
0.004^{+0.003}_{-0.001} \\
0.027^{+0.010}_{-0.009}
\end{array}$
&$\begin{array}{cc}
0.053^{+0.040}_{-0.016} \\
0.360^{+0.140}_{-0.118}
\end{array}$
\\
\hline
$R_{s, KK}^{\rm Theo}[f_{05}/\phi] \equiv
\frac{{\cal B}(B_s^0 \to J/\psi f_0(1500) (\to K^+ K^-))}
{{\cal B}(B_{s}^0 \to J/\psi \phi (\to K^{+} K^{-})) }$
&$\begin{array}{cc}
0.001^{+0.001}_{-0.000} \\
0.005^{+0.002}_{-0.002}
\end{array}$
&$\begin{array}{cc}
0.010^{+0.007}_{-0.003} \\
0.067^{+0.033}_{-0.023}
\end{array}$
\\
\hline \hline
\end{tabular}}
\end{center}
\end{table}

\hspace{0.5mm}
Following the experimental strategies, we define the interesting ratios by utilizing the
referenced channels $B_d^0 \to J/\psi \rho^0 (\to \pi^+ \pi^-)$ and $B_s^0 \to J/\psi
\phi (\to K^+ K^-)$ with ${\cal B}(\rho^0 \to \pi^+ \pi^-) \sim 100\%$ and
${\cal B}(\phi \to K^+ K^-)= 0.491 \pm 0.005$~\cite{ParticleDataGroup:2020ssz}.
The PQCD predictions about these ratios between the branching fractions of
$B^0 \to J/\psi f_0(1500)(\to \pi^+ \pi^-/K^+ K^-)$ and $B^0 \to J/\psi \rho^0/\phi
(\to \pi^+ \pi^-/K^+ K^-)$ are presented in Table~\ref{tab:Rs-BR-15}.	
By employing Eqs.~(\ref{eq:psif15s-pp-ex-lhcb}),~(\ref{eq:psiphi-ex}), and~(\ref{eq:psif15s-kk-ex}),
we can derive the relative ratios $R_{s, \pi K}^{\rm Exp}[f_{05}/\phi]$
and $R_{s, KK}^{\rm Exp}[f_{05}/\phi]$ from the experiment side as follows,
\beq
R_{s, \pi K}^{\rm Exp}[f_{05}/\phi] &\equiv&
\frac{{\cal B}(B_s^0 \to J/\psi f_0(1500) (\to \pi^+ \pi^-))_{\rm Exp}}
{{\cal B}(B_s^0 \to J/\psi \phi (\to K^+ K^-))_{\rm Exp}}
\non
&\simeq&
\frac{{\cal B}(B_s^0 \to J/\psi f_0(1500))\cdot {\cal B}(f_0(1500) \to \pi^+ \pi^-)}
{{\cal B}(B_s^0 \to J/\psi \phi)\cdot {\cal B}(\phi \to K^+ K^-)}
= 0.039^{+0.002}_{-0.002}\;,
\label{eq:r-pik-s-15-ex}\\
R_{s, KK}^{\rm Exp}[f_{05}/\phi] &\equiv&
\frac{{\cal B}(B_s^0 \to J/\psi f_0(1500) (\to K^+ K^-))_{\rm Exp}}
{{\cal B}(B_s^0 \to J/\psi \phi (\to K^+ K^-))_{\rm Exp}}
\non
&\simeq&
\frac{{\cal B}(B_s^0 \to J/\psi f_0(1500))\cdot {\cal B}(f_0(1500) \to K^+ K^-)}
{{\cal B}(B_s^0 \to J/\psi \phi)\cdot {\cal B}(\phi \to K^+ K^-)}
= 0.007^{+0.001}_{-0.001} \;.
\label{eq:r-kk-s-15-ex}
\eeq
It seems that the theoretical ratios $R_{s, \pi K}^{\rm Theo}[f_{05}/\phi]$
and $R_{s, KK}^{\rm Theo}[f_{05}/\phi]$ in the schemes of S2 with $M_{\rm I}$ and S1 with $M_{\rm II}$
are consistent with those derived from the experimental data within uncertainties.
It means that the present predictions in theory associated with the limited data in experiments
are not enough for us to identify the favorite model in the $N$-$S$-$G$ mixing.

\item[(3)]{The decays $B^0 \to J/\psi f_0(1710)$}

\hspace{0.5mm}
From the numerical results collected in Table~\ref{tab:BR-MI&II}, the large branching
fractions of the decays $B^0 \to J/\psi f_0(1710)$ within large uncertainties
could be read as,
\beq
{\cal B}(B_d^0 \to J/\psi f_0(1710))_{\rm S1} &=&
\biggl\{\begin{array}{ll}
(2.93^{+2.67}_{-1.29}) \times 10^{-7} \qquad  \ (M_{\rm I}) \\
(6.73^{+5.78}_{-2.99}) \times 10^{-7} \qquad  \ (M_{\rm II})
\end{array} \;,
\label{eq:psif17-d-s1}\\
{\cal B}(B_d^0 \to J/\psi f_0(1710))_{\rm S2} &=&
\biggl\{\begin{array}{ll}
(1.26^{+0.74}_{-0.65}) \times 10^{-6} \qquad  \ (M_{\rm I}) \\
(2.84^{+1.70}_{-1.44}) \times 10^{-6} \qquad  \ (M_{\rm II})
\end{array} \;,
\label{eq:psif17-d-s2}
\eeq
and
\beq
{\cal B}(B_s^0 \to J/\psi f_0(1710))_{\rm S1} &=&
\biggl\{\begin{array}{ll}
(1.36^{+1.31}_{-0.69}) \times 10^{-4} \qquad  \ (M_{\rm I}) \\
(1.07^{+1.09}_{-0.53}) \times 10^{-5} \qquad  \ (M_{\rm II})
\end{array} \;,
\label{eq:psif17-s-s1}\\
{\cal B}(B_s^0 \to J/\psi f_0(1710))_{\rm S2} &=&
\biggl\{\begin{array}{ll}
(9.10^{+5.81}_{-4.75}) \times 10^{-4} \qquad  \ (M_{\rm I}) \\
(7.32^{+4.68}_{-3.76}) \times 10^{-5} \qquad  \ (M_{\rm II})
\end{array} \;,
\label{eq:psif17-s-s2}
\eeq
where the errors from various sources have been added in quadrature.
Unfortunately, the decays $B^0 \to J/\psi f_0(1710)$ have not been observed
at any experiments yet. We therefore expect the near future measurements
on these branching fractions around $10^{-5} \sim 10^{-4}$
at Belle-II and LHCb experiments. The ratios between the branching fractions of
$B^0 \to J/\psi f_0(1710)$ and $B^0 \to J/\psi \rho^0/\phi$ are defined as,
\beq
R_d^{\rm Theo}[f_{07}/\rho]_{\rm S1}
&\equiv&
\frac{{\cal B}(B_d^0 \to J/\psi f_0(1710))_{\rm S1}}
{{\cal B}(B_d^0 \to J/\psi \rho^0)}
=
\biggl\{\begin{array}{ll}
0.010^{+0.007}_{-0.002}  \qquad  \ (M_{\rm I}) \\
0.023^{+0.015}_{-0.006}  \qquad  \ (M_{\rm II})
\end{array} \;,
\label{eq:f17rho-d-s1}
\\
R_d^{\rm Theo}[f_{07}/\rho]_{\rm S2}
&\equiv&
\frac{{\cal B}(B_d^0 \to J/\psi f_0(1710))_{\rm S2}}
{{\cal B}(B_d^0 \to J/\psi \rho^0)}
=
\biggl\{\begin{array}{ll}
0.042^{+0.015}_{-0.015} \qquad  \ (M_{\rm I}) \\
0.095^{+0.035}_{-0.033} \qquad  \ (M_{\rm II})
\end{array} \;,
\label{eq:f17rho-d-s2}
\eeq
and
\beq
R_s^{\rm Theo}[f_{07}/\phi]_{\rm S1}
&\equiv&
\frac{{\cal B}(B_s^0 \to J/\psi f_0(1710))_{\rm S1}}
{{\cal B}(B_s^0 \to J/\psi \phi)}
=
\biggl\{\begin{array}{ll}
0.127^{+0.087}_{-0.038}  \qquad  \ (M_{\rm I}) \\
0.010^{+0.008}_{-0.003}  \qquad  \ (M_{\rm II})
\end{array} \;,
\label{eq:f17phi-s-s1}
\\
R_s^{\rm Theo}[f_{07}/\phi]_{\rm S2}
&\equiv&
\frac{{\cal B}(B_s^0 \to J/\psi f_0(1710))_{\rm S2}}
{{\cal B}(B_s^0 \to J/\psi \phi)}
=
\biggl\{\begin{array}{ll}
0.851^{+0.306}_{-0.277} \qquad  \ (M_{\rm I}) \\
0.068^{+0.025}_{-0.022} \qquad  \ (M_{\rm II})
\end{array} \;.
\label{eq:f17phi-s-s2}
\eeq
Attributed to the dominance of
scalar $S (G)$ in the $f_0(1710)$ state in model I (II),
the experimental measurements on the evidently
large ratios predicted in the PQCD approach
would provide useful information to help identify the two
possible mixing models.

\begin{table}[htb]
\caption{ Same as Table~\ref{tab:BR4psif15-NWA} but for $B^0 \to J/\psi f_0(1710)$. }
\label{tab:BR4psif17-NWA}
\begin{center}\vspace{-0.5cm}{
\begin{tabular}[t]{c|c|c}
\hline  \hline
Decay modes   &  model I  & model II \\
\hline \hline
$B_d^0 \to J/\psi f_0(1710)(\to \pi^+ \pi^-)$
&$\begin{array}{cc}
(1.56^{+1.73}_{-1.04})
\times 10^{-8}
\\
(6.72^{+5.77}_{-4.83})
\times 10^{-8}
\end{array}$
&$\begin{array}{cc}
(3.59^{+3.81}_{-2.40})
\times 10^{-8}
\\
(1.51^{+1.32}_{-1.08})
\times 10^{-7}
\end{array}$
\\
\hline
$B_s^0 \to J/\psi f_0(1710)(\to \pi^+ \pi^-)$
&$\begin{array}{cc}
(7.25^{+8.33}_{-5.16})\times 10^{-6}
\\
(4.85^{+4.33}_{-3.51})\times 10^{-5}
\end{array}$
&$\begin{array}{cc}
(5.71^{+6.84}_{-4.02})
\times 10^{-7}
\\
(3.90^{+3.49}_{-2.80})\times 10^{-6}
\end{array}$
\\
\hline  \hline
$B_d^0 \to J/\psi f_0(1710)(\to K^+ K^-)$
&$\begin{array}{cc}
(5.27^{+5.11}_{-2.91})\times 10^{-8}
\\
(2.27^{+1.54}_{-1.39})
\times 10^{-7}
\end{array}$
&$\begin{array}{cc}
(1.21^{+1.11}_{-0.67})
\times 10^{-7}
\\
(5.11^{+3.50}_{-3.10})\times 10^{-7}
\end{array}$
\\
\hline
$B_s^0 \to J/\psi f_0(1710)(\to K^+ K^-)$
&$\begin{array}{cc}
(2.45^{+2.50}_{-1.49})
\times 10^{-5}
\\
(1.64^{+1.18}_{-1.02})
\times 10^{-4}
\end{array}$
&$\begin{array}{cc}
(1.93^{+2.07}_{-1.15})\times 10^{-6}
\\
(1.32^{+0.95}_{-0.81})\times 10^{-5}
\end{array}$
\\
\hline \hline
	\end{tabular}}
	\end{center}
\end{table}

\hspace{0.5mm}
As suggested in~\cite{Close:2015rza}, the resonance $f_0(1710)$ could be measured
in the $B_s^0$ decays in both the $\pi^+ \pi^-$ and $K^+ K^-$ channels at the LHCb experiments.
According to the strong decays of $f_0(1710)$, i.e., ${\cal B}(f_0(1710) \to K\bar{K})=0.36 \pm 0.12$ and
$\Gamma(f_0(1710) \to \pi\pi)/\Gamma(f_0(1710) \to K\bar{K})=0.23 \pm 0.05$~\cite{ParticleDataGroup:2020ssz},
${\cal B}(f_0(1710) \to \pi^+ \pi^-)= \frac{2}{3}(0.08^{+0.05}_{-0.04})$
and ${\cal B}(f_0(1710) \to K^+K^-)= 0.18 \pm 0.06$ could also be easily derived based on isospin symmetry.
The reliability of NWA for $B_s^0 \to J/\psi f_0(1710)$ has been discussed in~\cite{Cheng:2015iaa}.
Therefore, the large PQCD predictions for ${\cal B}(B_s^0 \to J/\psi f_0(1710) (\to \pi^+ \pi^-))$
and ${\cal B}(B_s^0 \to J/\psi f_0(1710) (\to K^+ K^-))$ are obtained under NWA.
Of course, when the appropriate two-pion and two-kaon distribution amplitudes are available,
they could also be investigated through the quasi-two-body $B$-meson decays.
The numerical results for ${\cal B}(B_s^0 \to J/\psi f_0(1710) (\to \pi^+ \pi^-/ K^+ K^-))$ in PQCD approach are presented in Table~\ref{tab:BR4psif17-NWA} with two different models and scenarios for $f_0$.

\begin{table}[htb]
\caption{ Same as Table~\ref{tab:Rs-BR-15} but for $B^0 \to J/\psi f_0(1710)$. }
\label{tab:Rs-BR-17}
\begin{center}\vspace{-0.5cm}{
\begin{tabular}[t]{c|c|c}
\hline  \hline
Ratios   &  model I  & model II \\
\hline \hline
$R_{d, \pi\pi}^{\rm Theo}[f_{07}/\rho] \equiv
\frac{{\cal B}(B_d^0 \to J/\psi f_0(1710) (\to \pi^+ \pi^-))}
{{\cal B}(B_{d}^0 \to J/\psi \rho^0 (\to \pi^{+} \pi^{-}))}$
&$\begin{array}{cc}
0.0005^{+0.0005}_{-0.0003} \\
0.002^{+0.002}_{-0.001}
\end{array}$
&$\begin{array}{cc}
0.001^{+0.001}_{-0.000} \\
0.005^{+0.002}_{-0.002}
\end{array}$
\\
\hline
$R_{d, K\pi}^{\rm Theo}[f_{07}/\rho] \equiv
\frac{{\cal B}(B_d^0 \to J/\psi f_0(1710) (\to K^+ K^-))}
{{\cal B}(B_{d}^0 \to J/\psi \rho^0 (\to \pi^{+} \pi^{-}))}$
&$\begin{array}{cc}
0.002^{+0.001}_{-0.001} \\
0.008^{+0.004}_{-0.004}
\end{array}$
&$\begin{array}{cc}
0.004^{+0.003}_{-0.001} \\
0.017^{+0.006}_{-0.006}
\end{array}$
\\
\hline
$R_{s, \pi K}^{\rm Theo}[f_{07}/\phi] \equiv
\frac{{\cal B}(B_s^0 \to J/\psi f_0(1710) (\to \pi^+ \pi^-))}
{{\cal B}(B_{s}^0 \to J/\psi \phi (\to K^{+} K^{-})) }$
&$\begin{array}{cc}
0.014^{+0.013}_{-0.008} \\
0.092^{+0.065}_{-0.054}
\end{array}$
&$\begin{array}{cc}
0.001^{+0.001}_{-0.001} \\
0.007^{+0.005}_{-0.004}
\end{array}$
\\
\hline
$R_{s, KK}^{\rm Theo}[f_{07}/\phi] \equiv
\frac{{\cal B}(B_s^0 \to J/\psi f_0(1710) (\to K^+ K^-))}
{{\cal B}(B_{s}^0 \to J/\psi \phi (\to K^{+} K^{-}))}$
&$\begin{array}{cc}
0.046^{+0.036}_{-0.020} \\
0.309^{+0.152}_{-0.143}
\end{array}$
&$\begin{array}{cc}
0.004^{+0.003}_{-0.002} \\
0.025^{+0.012}_{-0.011}
\end{array}$
\\
\hline \hline
\end{tabular}}
\end{center}
\end{table}

\hspace{0.5mm}
Similarly, we also predict the ratios between the branching fractions of $B^0 \to J/\psi f_0(1710) (\to \pi^+ \pi^-/ K^+ K^-)$
and $B^0 \to J/\psi \rho^0/\phi (\to \pi^+ \pi^-/ K^+ K^-)$. The numerical results
in PQCD approach collected in Table~\ref{tab:Rs-BR-17} await relevant tests in the future.
\end{itemize}

Additionally, according to the results in Table~\ref{tab:BR-MI&II} and the mixing coefficients in Eqs.~(\ref{eq:MI})
and~(\ref{eq:MII}), we can extract the PQCD branching fractions of $B_d^0 \to J/\psi f_0(N)$ and $B_s^0 \to J/\psi f_0(S)$
from $B_d^0 \to J/\psi f_0(1370)$ and $B_s^0 \to J/\psi f_0(1710)$ in $M_{\rm I}$
and from $B_d^0 \to J/\psi f_0(1370)$ and $B_s^0 \to f_0(1500)$ in $M_{\rm II}$, respectively, as follows,
\begin{itemize}
\item {In model I:}
\beq
{\cal B}(B_d^0 \to J/\psi f_0(N))_{M_{\rm I}} &=&
\biggl\{\begin{array}{ll}
(1.30^{+1.14}_{-0.54}) \times 10^{-5} \qquad    ({\rm S1})\\
(5.56^{+3.03}_{-2.69}) \times 10^{-5}\qquad   
({\rm S2})
\end{array}\;,
\label{eq:BR-psiN-I}\\
{\cal B}(B_s^0 \to J/\psi f_0(S))_{M_{\rm I}} &=&
\biggl\{\begin{array}{ll}
(1.53^{+1.42}_{-0.73}) \times 10^{-4} \qquad    ({\rm S1})\\
(1.03^{+0.60}_{-0.52}) \times 10^{-3} \qquad   
({\rm S2})
\end{array}\;,
\label{eq:BR-psiS-I}
\eeq
		
\item {In model II:}
\beq
{\cal B}(B_d^0 \to J/\psi f_0(N))_{M_{\rm II}} &=&
\biggl\{\begin{array}{ll}
(1.40^{+1.14}_{-0.59}) \times 10^{-5} \qquad    ({\rm S1})\\
(5.89^{+3.19}_{-2.84}) \times 10^{-5} \qquad    ({\rm S2})
\end{array}\;,
\label{eq:BR-psiN-II}\\
{\cal B}(B_s^0 \to J/\psi f_0(S))_{M_{\rm II}} &=&
\biggl\{\begin{array}{ll}
(1.62^{+1.55}_{-0.77}) \times 10^{-4} \qquad   ({\rm S1})\\
(1.10^{+0.64}_{-0.55}) \times 10^{-3} \qquad   ({\rm S2})
\end{array}\;.
\label{eq:BR-psiS-II}
\eeq
\end{itemize}
Here, the consistent while slightly different values for the branching fractions of
$B_d^0 \to J/\psi f_0(N)$ and $B_s^0 \to J/\psi f_0(S)$
are actually induced by the slightly different masses for scalar
quarkonia $N$ and $S$ in the two different mixing models I and II.
These branching fractions are basically consistent with those in
Eqs.~(\ref{eq:psif0-nn-NF}) and~(\ref{eq:psif0-ss-NF})
by using naive factorization~\cite{Wang:2009cb,Wang:2009rc} within dramatically
large uncertainties.
But, it is also clear to see that our predictions are a bit smaller than
those in Eqs.~(\ref{eq:psif0-nn-NF}) and~(\ref{eq:psif0-ss-NF}) explicitly.
It implies that there might have large nonfactorizable contributions
in these types of color-suppressed-tree-dominated $B$-meson decays~\cite{Chen:2005ht}
to destructively interfere with the factorizable emission amplitudes.

In principle, according to Eqs.~(\ref{eq:DecAmp-psif0-d}) and (\ref{eq:DecAmp-psif0-s}),
the $B_{d(s)}^0 \to J/\psi f_0$ branching fractions could be obtained straightforwardly
through multiplying $|\alpha_1|^2 (|\alpha_2|^2)$ by Eqs.~(\ref{eq:BR-psiN-I}) and
(\ref{eq:BR-psiN-II}) [(\ref{eq:BR-psiS-I}) and~(\ref{eq:BR-psiS-II})].
Thus, in other words, with the help of Eqs.~(\ref{eq:BR-psiN-I})-(\ref{eq:BR-psiS-II}) and the
measurements on the decays $B^0 \to J/\psi f_0$, one could directly determine
the coefficients $|\alpha_1|^2$ and $|\alpha_2|^2$ in the PQCD approach,
though suffering from large uncertainties, which could further
give the information about the amount of scalar glueball components by
Eq.~(\ref{eq:normalization}). Unfortunately, quite few measurements on the decays
$B^0 \to J/\psi f_0$, besides the available data of
${\cal B}(B_s^0 \to J/\psi f_0(1370, 1500), f_0 \to \pi^+ \pi^-)$,
are available currently, which limits our detailed studies on
these scalars $f_0$.

\begin{table}[!ht]
\caption{
Relative ratios of the branching fractions between the different $B^0 \to J/\psi f_0$ decays.
The upper (lower) entry corresponds to $f_0$ in scenario 1 (2) at every line.}
\label{tab:Rs-MI&II}
\begin{center}\vspace{-0.5cm}{
\begin{tabular}[t]{c|c|c}
\hline  \hline
Ratios   &  model I & model II \\
\hline \hline
$R_{d}^{\rm Theo}[f_{03}/f_{05}]\equiv
\frac{{\cal B}(B_{d}^{0} \to J/\psi f_{0}(1370))}
{{\cal B}(B_{d}^{0} \to J/\psi f_{0}(1500))}$
&
$\hspace{1.8mm}
2.41^{+0.00}_{-0.01}
$
&
$\hspace{1.8mm}
4.96^{+0.01}_{-0.01}
$
\\
\hline
$R_{d}^{\rm Theo}[f_{03}/f_{07}]\equiv
\frac{{\cal B}(B_{d}^{0} \to J/\psi f_{0}(1370))}
{{\cal B}(B_{d}^{0} \to J/\psi f_{0}(1710))}$
&
$
29.11^{+0.04}_{-0.03}
$
&
$
15.44^{+0.05}_{-0.05}
$
\\
\hline
$R_{d}^{\rm Theo}[f_{05}/f_{07}]\equiv
\frac{{\cal B}(B_{d}^{0} \to J/\psi f_{0}(1500))}
{{\cal B}(B_{d}^{0} \to J/\psi f_{0}(1710))}$
&
$
12.10^{+0.04}_{-0.00}
$
&
$\hspace{1.8mm}
3.11^{+0.02}_{-0.01}
$
\\
\hline
$R_{s}^{\rm Theo}[f_{05}/f_{03}]\equiv
\frac{{\cal B}(B_{s}^{0} \to J/\psi f_{0}(1500))}
{{\cal B}(B_{s}^{0} \to J/\psi f_{0}(1370))}$
&
$\hspace{1.8mm}
1.83^{+0.01}_{-0.01}
$
&
$\hspace{1.8mm}
6.40^{+0.01}_{-0.01}
$
\\
\hline
$R_{s}^{\rm Theo}[f_{07}/f_{03}]\equiv
\frac{{\cal B}(B_{s}^{0} \to J/\psi f_{0}(1710))}
{{\cal B}(B_{s}^{0} \to J/\psi f_{0}(1370))}$
&
$
26.93^{+0.11}_{-0.13}
$
&
$\hspace{1.8mm}
0.56^{+0.00}_{-0.00}
$
\\
\hline
$R_{s}^{\rm Theo}[f_{05}/f_{07}]\equiv
\frac{{\cal B}(B_{s}^{0} \to J/\psi f_{0}(1500))}
{{\cal B}(B_{s}^{0} \to J/\psi f_{0}(1710))}$
&
$\hspace{1.8mm}
0.07^{+0.00}_{-0.00}
$
&
$
11.35^{+0.04}_{-0.04}
$
\\
\hline \hline
$R_{s/d}^{\rm Theo}[f_{03}] \equiv
\frac{{\cal B}(B_{s}^{0} \to J/\psi f_{0}(1370))}
{{\cal B}(B_{d}^{0} \to J/\psi f_{0}(1370))}$
&
$\begin{array}{cc}
0.59^{+0.08}_{-0.08}
\\
0.93^{+0.08}_{-0.06}
\end{array}$
&
$\begin{array}{cc}
\hspace{1.8mm}
1.83^{+0.24}_{-0.18}
\\
\hspace{1.8mm}
2.97^{+0.23}_{-0.15}
\end{array}$
\\
\hline
$R_{s/d}^{\rm Theo}[f_{05}] \equiv
\frac{{\cal B}(B_{s}^{0} \to J/\psi f_{0}(1500))}
{{\cal B}(B_{d}^{0} \to J/\psi f_{0}(1500))}$
&
$\begin{array}{cc}
2.61^{+0.35}_{-0.34}
\\
4.06^{+0.33}_{-0.26}
\end{array}$
&
$\begin{array}{cc}
58.10^{+7.03}_{-6.23}
\\
94.12^{+6.96}_{-4.31}
\end{array}$
\\
\hline
$R_{s/d}^{\rm Theo}[f_{07}]\equiv
\frac{{\cal B}(B_{s}^{0} \to J/\psi f_{0}(1710))}
{{\cal B}(B_{d}^{0} \to J/\psi f_{0}(1710))}$
&
$\begin{array}{cc}
(4.64^{+0.62}_{-0.60})\times 10^{2}
\\
(7.22^{+0.66}_{-0.39})\times 10^{2}
\end{array}$
&
$\begin{array}{cc}
15.90^{+2.07}_{-1.60}
\\
25.77^{+1.99}_{-1.20}
\end{array}$
\\
\hline \hline
$R_{s/d}^{\rm Theo}[f_{05}/f_{03}] \equiv
\frac{{\cal B}(B_{s}^{0} \to J/\psi f_{0}(1500))}
{{\cal B}(B_{d}^{0} \to J/\psi f_{0}(1370))}$
&
$\begin{array}{cc}
\hspace{1.8mm}
1.08^{+0.14}_{-0.14}
\\
\hspace{1.8mm}
1.69^{+0.13}_{-0.12}
\end{array}$
&
$\begin{array}{cc}
11.73^{+1.42}_{-1.23}
\\
19.00^{+1.38}_{-0.91}
\end{array}$
\\
\hline
$R_{s/d}^{\rm Theo}[f_{07}/f_{03}] \equiv
\frac{{\cal B}(B_{s}^{0} \to J/\psi f_{0}(1710))}
{{\cal B}(B_{d}^{0} \to J/\psi f_{0}(1370))}$
&
$\begin{array}{cc}
15.93^{+2.15}_{-2.04}
\\
24.93^{+1.94}_{-1.70}
\end{array}$
&
$\begin{array}{cc}
\hspace{1.8mm}
1.03^{+0.14}_{-0.10}
\\
\hspace{1.8mm}
1.67^{+0.13}_{-0.08}
\end{array}$
\\
\hline
$R_{s/d}^{\rm Theo}[f_{07}/f_{05}] \equiv
\frac{{\cal B}(B_{s}^{0} \to J/\psi f_{0}(1710))}
{{\cal B}(B_{d}^{0} \to J/\psi f_{0}(1500))}$
&
$\begin{array}{cc}
38.31^{+5.16}_{-4.92}
\\
59.87^{+4.92}_{-3.85}
\end{array}$
&
$\begin{array}{cc}
\hspace{1.8mm}
5.10^{+0.69}_{-0.51}
\\
\hspace{1.8mm}
8.28^{+0.64}_{-0.38}
\end{array}$
\\
\hline
$R_{s/d}^{\rm Theo}[f_{03}/f_{05}] \equiv
\frac{{\cal B}(B_{s}^{0} \to J/\psi f_{0}(1370))}
{{\cal B}(B_{d}^{0} \to J/\psi f_{0}(1500))}$
&
$\begin{array}{cc}
\hspace{1.8mm}
1.43^{+0.19}_{-0.19}
\\
\hspace{1.8mm}
2.22^{+0.18}_{-0.15}
\end{array}$
&
$\begin{array}{cc}
\hspace{1.8mm}
9.05^{+1.17}_{-0.89}
\\
14.71^{+1.15}_{-0.71}
\end{array}$
\\
\hline
$R_{s/d}^{\rm Theo}[f_{03}/f_{07}] \equiv
\frac{{\cal B}(B_{s}^{0} \to J/\psi f_{0}(1370))}
{{\cal B}(B_{d}^{0} \to J/\psi f_{0}(1710))}$
&
$\begin{array}{cc}
17.30^{+2.30}_{-2.27}
\\
26.83^{+2.43}_{-1.52}
\end{array}$
&
$\begin{array}{cc}
28.23^{+3.54}_{-2.84}
\\
45.77^{+3.59}_{-2.29}
\end{array}$
\\
\hline
$R_{s/d}^{\rm Theo}[f_{05}/f_{07}] \equiv
\frac{{\cal B}(B_{s}^{0} \to J/\psi f_{0}(1500))}
{{\cal B}(B_{d}^{0} \to J/\psi f_{0}(1710))}$
&
$\begin{array}{cc}
(0.32^{+0.04}_{-0.04}) \times 10^2
\\
(0.49^{+0.04}_{-0.03}) \times 10^2
\end{array}$
&
$\begin{array}{cc}
(1.81^{+0.21}_{-0.20}) \times 10^2
 \\
(2.93^{+0.22}_{-0.14}) \times 10^2
\end{array}$
\\
\hline \hline
\end{tabular}}
\end{center}
\end{table}

Undoubtedly, the above listed branching fractions associated with their relative ratios suffer
from large theoretical errors coming from various hadronic parameters.
While, generally speaking, the theoretical uncertainties resulted from
the same hadronic inputs could be cancelled in the ratios to a great extent.
Therefore, based on the results shown in Table~\ref{tab:BR-MI&II},
we define several ratios of those PQCD branching fractions and present their values
in Table~\ref{tab:Rs-MI&II}, where we find that the uncertainties from hadronic
parameters could be cancelled greatly in almost all of the ratios.
These ratios could be naively classified into three types.
\begin{itemize}
\item{Type-1: the ratios between the branching fractions of the two modes containing different
final states but with same transition amplitudes }

For example, as displayed in the first six lines of Table~\ref{tab:Rs-MI&II},
the ratios between the branching fractions of $B^0 \to J/\psi f_{0}(1500)$
and $B^0 \to J/\psi f_{0}(1710)$ could be analytically written as,
\beq
R_{d}^{\rm Theo}[f_{05}/f_{07}]
&\equiv&
\frac{{\cal B}(B_{d}^{0} \to J/\psi f_{0}(1500))}
{{\cal B}(B_{d}^{0} \to J/\psi f_{0}(1710))}
=
\frac{\Phi(r^d_{J/\psi}, r^d_{f_{05}})}
{\Phi(r^d_{J/\psi}, r^d_{f_{07}})}
\cdot
\frac{|\alpha_{1}^{f_{05}}|^2}{|\alpha_{1}^{f_{07}}|^2}
\;,
\label{eq:rd-15ov17}
\\
R_{s}^{\rm Theo}[f_{05}/f_{07}]
&\equiv&
\frac{{\cal B}(B_{s}^{0} \to J/\psi f_{0}(1500))}
{{\cal B}(B_{s}^{0} \to J/\psi f_{0}(1710))}
=
\frac{\Phi(r^s_{J/\psi}, r^s_{f_{05}})}
{\Phi(r^s_{J/\psi}, r^s_{f_{07}})}
\cdot
\frac{|\alpha_{2}^{f_{05}}|^2}{|\alpha_{2}^{f_{07}}|^2}
\;,
\label{eq:rs-15ov17}
\eeq
which can give the information about the mixing coefficients $\alpha_1$ and $\alpha_2$ of $f_{0}(1500)$
and $f_{0}(1710)$ in the related channels cleanly. Furthermore, as exhibited in Table~\ref{tab:Rs-MI&II},
the values of these ratios with highly evident deviation could help us
differentiate the possible mixing model when the precise measurements are
available in the near future. From these two representative ratios,
we can expect that these kinds of scenario-independent ratios could be utilized to explore
the relations of the coefficients in the $N$-$S$-$G$ mixing.

\item{Type-2: the ratios between the branching fractions of the two modes containing same
final states but with different transition amplitudes }

For example, as presented in the second three lines of Table~\ref{tab:Rs-MI&II},
the ratios between the branching fractions of $B_s^0 \to J/\psi f_{0}$
and $B_d^0 \to J/\psi f_{0}$ could be analytically expressed as,
\beq
R_{s/d}^{\rm Theo}[f_{05}]
&\equiv&
\frac{{\cal B}(B_{s}^{0} \to J/\psi f_{0}(1500))}
{{\cal B}(B_{d}^{0} \to J/\psi f_{0}(1500))}
\non
&=&
\frac{\tau_{B_s^0}}{\tau_{B_d^0}}
\cdot
\frac{\Phi(r^s_{J/\psi}, r^s_{f_{05}})}
{\Phi(r^d_{J/\psi}, r^d_{f_{05}})}
\cdot
\frac{|A(B_s^0 \to J/\psi f_0(S))|^2}{|A(B_d^0 \to J/\psi f_0(N)|^2}
\cdot
\frac{|\alpha_{2}^{f_{05}}|^2}{|\alpha_{1}^{f_{05}}|^2}
\;,
\label{eq:rsd-15ov15}
\eeq
\\
\beq
R_{s/d}^{\rm Theo}[f_{07}]
&\equiv&
\frac{{\cal B}(B_{s}^{0} \to J/\psi f_{0}(1710))}
{{\cal B}(B_{d}^{0} \to J/\psi f_{0}(1710))}
\non
&=&
\frac{\tau_{B_s^0}}{\tau_{B_d^0}}
\cdot
\frac{\Phi(r^s_{J/\psi}, r^s_{f_{07}})}
{\Phi(r^d_{J/\psi}, r^d_{f_{07}})}
\cdot
\frac{|A(B_s^0 \to J/\psi f_0(S))|^2}{|A(B_d^0 \to J/\psi f_0(N)|^2}
\cdot
\frac{|\alpha_{2}^{f_{07}}|^2}{|\alpha_{1}^{f_{07}}|^2}
\;.
\label{eq:rsd-17ov17}
\eeq
These ratios could tell us the relations about the coefficients $\alpha_1$ and $\alpha_2$
in the same scalar state in a relatively clean manner under the SU(3) limit. At the same time,
if the ratio $\alpha_2^{f_0}/\alpha_1^{f_0}$ could be determined precisely from the experimental data,
then the broken SU(3) flavor symmetry could also be explored in these related decays, which can be
further used to understand the QCD of $f_S$ and $f_N$ deeply.

\item{Type-3: the ratios between the branching fractions of the two modes containing different
final states while with different transition amplitudes }

For example, as exhibited in the last six lines of Table~\ref{tab:Rs-MI&II}, the ratios between the branching
fractions of $B_s^0 \to J/\psi f_{0}(1500)(f_{0}(1710))$ and $B_d^0 \to J/\psi f_{0}(1710)(f_{0}(1500))$ can be
analytically described as,
\beq
R_{s/d}^{\rm Theo}[f_{05}/f_{07}]
&\equiv&
\frac{{\cal B}(B_{s}^{0} \to J/\psi f_{0}(1500))}
{{\cal B}(B_{d}^{0} \to J/\psi f_{0}(1710))}
\non
&=&
\frac{\tau_{B_s^0}}{\tau_{B_d^0}}
\cdot
\frac{\Phi(r^s_{J/\psi}, r^s_{f_{05}})}
{\Phi(r^d_{J/\psi}, r^d_{f_{07}})}
\cdot
\frac{|A(B_s^0 \to J/\psi f_0(S))|^2}{|A(B_d^0 \to J/\psi f_0(N)|^2}
\cdot
\frac{|\alpha_{2}^{f_{05}}|^2}{|\alpha_{1}^{f_{07}}|^2}
\;,
\label{eq:rsd-15ov17}
\\
R_{s/d}^{\rm Theo}[f_{07}/f_{05}]
&\equiv&
\frac{{\cal B}(B_{s}^{0} \to J/\psi f_{0}(1710))}
{{\cal B}(B_{d}^{0} \to J/\psi f_{0}(1500))}
\non
&=&
\frac{\tau_{B_s^0}}{\tau_{B_d^0}}
\cdot
\frac{\Phi(r^s_{J/\psi}, r^s_{f_{07}})}
{\Phi(r^d_{J/\psi}, r^d_{f_{05}})}
\cdot
\frac{|A(B_s^0 \to J/\psi f_0(S))|^2}{|A(B_d^0 \to J/\psi f_0(N)|^2}
\cdot
\frac{|\alpha_{2}^{f_{07}}|^2}{|\alpha_{1}^{f_{05}}|^2}
\;.
\label{eq:rsd-17ov15}
\eeq
These theoretically large ratios with small uncertainties could be tested at the relevant experiments,
though involving the complicated entanglements of the SU(3) symmetry breaking effects and the
mixing coefficients. Particularly, the first four ratios in type-3 have evidently large discrepancies
in the considered two different models, which could be used to identify the favorite model that prefer
the potential scalar glueball tentatively.

\end{itemize}

As presented in Sect.~\ref{sect:1}, the decays $B^0 \to J/\psi f_0(1370,1710)$ have ever been studied
under the assumption of $f_0(1500)$ being primarily a scalar glueball (corresponding to model I in this work)
from the hadron physics side~\cite{Xie:2014gla}. Three ratios from related decay widths have also been obtained.
Therefore, we quote our PQCD predictions as shown in the second column of Table~\ref{tab:Rs-MI&II} to
make comparisons with these three sets of ratios as collected in Eq.~(\ref{eq:Rs-Xie}). And, we can find that,
for the first and last ratios in~\cite{Xie:2014gla}, the PQCD values $(29.11^{+0.04}_{-0.03})$
and $(3.71^{+0.02}_{-0.01}) \times 10^{-2}$ are larger than the ratios in Eq.~(\ref{eq:Rs-Xie})
with the factors around $3.7 \sim 6.3$ and $2.7 \sim 4.6$ correspondingly.
For the second ratio, the result in Eq.~(\ref{eq:Rs-Xie}) is larger than our PQCD values
$(2.16^{+0.32}_{-0.26})\times 10^{-3}$ in scenario 1 and $(1.39^{+0.08}_{-0.12}) \times 10^{-3}$ in scenario 2
with the factors about $3.1 \sim 3.9$ and $4.6 \sim 6.5$, respectively.
These evident discrepancies would be confronted with the future examinations
and could further infer the amount of the scalar glueball components in the $f_0$ qualitatively.	

All the above predictions in the PQCD approach are expected to be measured at LHCb, Belle-II,
even the proposed CEPC experiments in the future,
and they could further provide more helpful constraints on discriminating the real scalar glueball.

%
%
\section{Conclusions and Summary}
\label{sect:4}
We have systematically calculated the branching fractions of the
decays $B^0 \to J/\psi f_0$ in
PQCD approach and $B^0 \to J/\psi f_0 (f_0 \to \pi^+ \pi^-/K^+ K^-)$
under narrow-width approximation by including the
vertex corrections at NLO level. Here, the light scalars $f_0$ include $f_0(1370)$,
$f_0(1500)$, and $f_0(1710)$, respectively. With two different
scenarios for $f_0$, the predictions of large branching fractions around $10^{-6} \sim 10^{-4}$
for $B^0 \to J/\psi f_0$ are obtained, and the results
await the future precise examinations at experiments.
According to the numerical results and phenomenological analyses, we conclude that:
\begin{itemize}
\item[]{(a)}
Large branching fractions mainly around $10^{-5}$ in both $B_d^0 \to J/\psi f_0(1370)$ and
$B_s^0 \to J/\psi f_0(1710)$ modes could be tested
at the LHCb and Belle-II experiments. The measurements with high precision can help us
understand the nature of $f_0(1370)$ and $f_0(1710)$.

\item[]{(b)}
The PQCD predictions about the branching fraction of $B_s^0 \to J/\psi f_0(1500)(\to \pi^+ \pi^-)$
in the schemes of S1 with $M_{\rm II}$ and S2 with $M_{\rm I}$ are consistent with each other and
agree well with the currently available data within large theoretical uncertainties.
The predicted ${\cal B}(B_s^0 \to J/\psi f_0(1500) (\to K^+ K^-))$ around $10^{-6}$
could be tested in the future experiments.

\item[]{(c)}
The large branching fractions of the decay $B_s^0 \to J/\psi f_0(1710)$ predicted in
PQCD approach provide one more chance to directly explore the flavor $s\bar s$ content
in the $f_0(1710)$ state. The forthcoming measurements of
${\cal B}(B_s^0 \to J/\psi f_0(1500, 1710))$ could isolate the $s\bar s$ components
with different but definite coefficients in the possible candidates of scalar glueball.

\item[]{(d)}
Several interesting relative ratios of the branching fractions between the different $B^0 \to J/\psi f_0$ decays
such as $R_d^{\rm Theo}[f_{05}/f_{07}]$, $R_s^{\rm Theo}[f_{05}/f_{07}]$, $R_s^{\rm Theo}[f_{07}/f_{03}]$,
and so forth, can give the information about the mixing coefficients $\alpha_1$ and $\alpha_2$ in the light
scalars $f_0$ in a clean and complementary manner. Then the amount of the scalar glueball component in
$f_0(1500)$ and $f_0(1710)$ could be deduced. They would be utilized to conjecture which one is more
favored as the primary scalar glueball.
\end{itemize}
Frankly speaking, in order to figure out the real scalar glueball clearly, we are eagerly looking forward to the
more stringent calculations on the nonperturbative parameters
and the more precise measurements on the undetected modes in this work.
The measurements with good precision on the decays $B^0 \to J/\psi f_0(1500, 1710)$
would provide the great opportunities to the lightest scalar glueball hunting potentially.

%
%
	
\begin{acknowledgments}
J.R. thanks D.Y. for his helpful discussions.
This work is supported by the National Natural Science
Foundation of China under Grants Nos.~11875033, 11775117, 11705159 and~12375089,
by the Qing Lan Project of Jiangsu Province under Grant No.~9212218405,
by the Natural Science Foundation of Shandong province under the Grants
No.~ZR2019JQ04 and~ZR2022ZD26,
and by the Research Fund of Jiangsu Normal University under Grant No.~HB2016004.
J.R. is supported by Postgraduate Research $\&$ Practice Innovation Program
of Jiangsu Normal University under Grant No.~2021XKT1235.
\end{acknowledgments}

%
%
	
\begin{appendix}	
\section{\boldmath Mixing matrices for scalars $f_0$ }
\label{sec:app1}
		
In this appendix, we will collect the mixing matrices for the considered $f_0$ quoted in this work.
Firstly, for the mixing model I with $f_0(1500)$ being the possible scalar glueball,
\begin{itemize}
\item
In~\cite{Close:2000yk, Close:2001ga}, Close and Kirk obtained the mixing matrices
in favor of $f_0(1500)$ as a glueball mixing with possible large $q\bar{q}$ states.
We take the absolute values and average them to get the following mixing matrix as,
\beq
\left(\begin{array}{cccc}
\vert f_0(1370)\rangle \\
\vert f_0(1500)\rangle \\
\vert f_0(1710)\rangle \\
\end{array} \right )
=\left(\begin{array}{cccc}
0.730 & 0.150 & 0.660\\
0.685 & 0.350 & 0.645\\
0.155 & 0.900 & 0.405\\
\end{array} \right )\
\left(\begin{array}{cccc}
\vert N \rangle \\
\vert S \rangle \\
\vert G \rangle \\
\end{array} \right )
\eeq

\item		
In~\cite{Giacosa:2004ug}, Giacosa {\it et al}. utilized a covariant constituent approach to analyze
glueball-quarkonia mixing and obtained the mixing matrix.
\beq
\left(\begin{array}{cccc}
\vert f_0(1370)\rangle \\
\vert f_0(1500)\rangle \\
\vert f_0(1710)\rangle \\
\end{array} \right )
=\left(\begin{array}{cccc}
\hspace{3.2mm}0.80 & 0.10 & \hspace{3.2mm}0.59\\
-0.59 & 0.25 & \hspace{3.2mm}0.76\\
\hspace{3.2mm}0.07 & 0.96 & -0.27\\
\end{array} \right )\
\left(\begin{array}{cccc}
\vert N \rangle \\
\vert S \rangle \\
\vert G \rangle \\
\end{array} \right )
\eeq

\item		
In~\cite{Close:2005vf}, Close and Zhao proposed a factorization scheme for studying the production of
$f_0$ in the hadronic decays into the isoscalar vector meson and
pseudoscalar meson pairs. Their results highlight the strong possibility of the existence
of glueball contents in $f_0(1500)$ correlated with the mixing matrix,
\beq
\left(\begin{array}{cccc}
\vert f_0(1370)\rangle \\
\vert f_0(1500)\rangle \\
\vert f_0(1710)\rangle \\
\end{array} \right )
=\left(\begin{array}{cccc}
-0.91 & -0.07 & \hspace{3.2mm}0.40\\
-0.41 & \hspace{3.2mm}0.35 & -0.84\\
\hspace{3.2mm}0.09 & \hspace{3.2mm}0.93 & \hspace{3.2mm}0.36\\
\end{array} \right )\
\left(\begin{array}{cccc}
\vert N \rangle \\
\vert S \rangle \\
\vert G \rangle \\
\end{array} \right )
\eeq

\item
In~\cite{Giacosa:2005qr}, Gutsche {\it et al}. discussed the phenomenological consequences of
the scalar meson sector in the context of an effective chiral Lagrangian and extracted the
possible glueball-quarkonia mixing scenario.
\beq
\left(\begin{array}{cccc}
\vert f_0(1370)\rangle \\
\vert f_0(1500)\rangle \\
\vert f_0(1710)\rangle \\
\end{array} \right )
=\left(\begin{array}{cccc}
\hspace{3.2mm} 0.86 & \hspace{3.2mm} 0.24 & \hspace{3.2mm} 0.45\\
-0.45 & -0.06 & \hspace{3.2mm} 0.89\\
-0.24 & \hspace{3.2mm} 0.97 & -0.06\\
\end{array} \right )\
\left(\begin{array}{cccc}
\vert N \rangle \\
\vert S \rangle \\
\vert G \rangle \\
\end{array} \right )
\eeq
	
\item	
In~\cite{Chatzis:2011qz}, Chatzis {\it et al}. obtained
two possible solutions by using a phenomenological Lagrangian approach.
In the first solution, the bare glueball dominantly resided in the $f_0(1500)$,
\beq
\left(\begin{array}{cccc}
\vert f_0(1370)\rangle \\
\vert f_0(1500)\rangle \\
\vert f_0(1710)\rangle \\
\end{array} \right )
=\left(\begin{array}{cccc}
\hspace{3.2mm}0.75 & \hspace{3.2mm}0.26 & \hspace{3.2mm}0.60\\
-0.59 & -0.14 & \hspace{3.2mm}0.80\\
-0.29 & \hspace{3.2mm}0.95 & -0.05\\
\end{array} \right )\
\left(\begin{array}{cccc}
\vert N \rangle \\
\vert S \rangle \\
\vert G \rangle \\
\end{array} \right )
\eeq
\end{itemize}

While, for the mixing model II with $f_0(1710)$ being the possible scalar glueball,
\begin{itemize}
\item
And also in~\cite{Chatzis:2011qz}, the scalar $f_0(1710)$ containing the largest glueball
component was suggested in the second solution associated with the mixing matrix,
\beq
\left(\begin{array}{cccc}
\vert f_0(1370)\rangle \\
\vert f_0(1500)\rangle \\
\vert f_0(1710)\rangle \\
\end{array} \right )
=\left(\begin{array}{cccc}
0.79 & \hspace{2.8mm}0.55 & \hspace{2.8mm}0.29\\
0.57 & -0.82 & \sim 0\\
0.23 & \hspace{2.8mm}0.17 & -0.96\\
\end{array} \right )\
\left(\begin{array}{cccc}
\vert N \rangle \\
\vert S \rangle \\
\vert G \rangle \\
\end{array} \right )
\eeq
	
\item	
In~\cite{Li:2000yn}, Li discussed the glueball-quarkonia content of
the three $f_0$ states taking into the two possible assumptions $M_G > M_S > M_N$ and
$M_G > M_N > M_S$ and obtained two solutions for the mixing matrix. Here, we quote
the first result for consideration.
\beq
\left(\begin{array}{cccc}
\vert f_0(1370)\rangle \\
\vert f_0(1500)\rangle \\
\vert f_0(1710)\rangle \\
\end{array} \right )
=\left(\begin{array}{cccc}
-0.922 & -0.121 & \hspace{3.2mm}0.367\\
-0.266 & \hspace{3.2mm}0.885 & -0.383\\
\hspace{3.2mm}0.276 & \hspace{3.2mm}0.455 & \hspace{3.2mm}0.847\\
\end{array} \right )\
\left(\begin{array}{cccc}
\vert N \rangle \\
\vert S \rangle \\
\vert G \rangle \\
\end{array} \right )
\eeq

\item		
In~\cite{Janowski:2013uga,Janowski:2014ppa}, Janowski and Giacosa investigated
the masses and decays of the three scalar-isoscalar resonances $f_0(1370)$, $f_0(1500)$ and $f_0(1710)$
in the framework of the extended Linear Sigma Model. Only solutions in which $f_0(1710)$
is predominantly a glueball were found. Again, we take the absolute values and average them to
get the following mixing matrix,
\beq
\left(\begin{array}{cccc}
\vert f_0(1370)\rangle \\
\vert f_0(1500)\rangle \\
\vert f_0(1710)\rangle \\
\end{array} \right )
=\left(\begin{array}{cccc}
0.905 & 0.325 & 0.190\\
0.360 & 0.920 & 0.100\\
0.155 & 0.155 & 0.960\\
\end{array} \right )\
\left(\begin{array}{cccc}
\vert N \rangle \\
\vert S \rangle \\
\vert G \rangle \\
\end{array} \right )
\eeq

\item		
In~\cite{Cheng:2015iaa}, Cheng {\it et al}. updated their study in~\cite{Cheng:2006hu} and
presented their newest results of the mixing matrix as follows,
\beq
\left(\begin{array}{cccc}
\vert f_0(1370)\rangle \\
\vert f_0(1500)\rangle \\
\vert f_0(1710)\rangle \\
\end{array} \right )
=\left(\begin{array}{cccc}
\hspace{2.8mm}0.78\pm 0.02 & 0.52\pm 0.03 & -0.36\pm 0.01\\
-0.55\pm 0.03 & 0.84\pm 0.02 & \hspace{2.8mm}0.03\pm 0.02\\
\hspace{2.8mm}0.31\pm 0.01 & 0.17\pm 0.01 & 0.934\pm 0.004\\
\end{array} \right )\
\left(\begin{array}{cccc}
\vert N \rangle \\
\vert S \rangle \\
\vert G \rangle \\
\end{array} \right )
\eeq

\item		
In~\cite{Guo:2020akt}, Guo {\it et al}. made a phenomenological
study fully based on the available data and found that
in $f_0(1710)$ a glueball component dominates.
The values of the mixing matrix could be read as,
\beq
\left(\begin{array}{cccc}
\vert f_0(1370)\rangle \\
\vert f_0(1500)\rangle \\
\vert f_0(1710)\rangle \\
\end{array} \right )
=\left(\begin{array}{cccc}
-0.96\sim-0.87 & -0.21\sim-0.07 & -0.45\sim-0.25\\
0.14\sim0.41 & -0.94\sim-0.82 & -0.40\sim-0.30\\
-0.36\sim-0.17 & -0.53\sim-0.32 & 0.80\sim0.92\\
\end{array} \right )\
\left(\begin{array}{cccc}
\vert N \rangle \\
\vert S \rangle \\
\vert G \rangle \\
\end{array} \right )
\eeq
\end{itemize}
		

\section{Related functions in PQCD approach}
\label{sec:app3}
		
The hard functions $h_i$ in the decay amplitudes come from the Fourier transformations
of the hard kernel, $H^{(0)}$ are as follows,
\beq
h_{fe}(x_{1},x_{3},b_{1},b_{3})= && K_{0}\biggl (\sqrt{x_{1}x_{3}(1-r_{2}^{2})}m_{B}b_{1}\biggl)
\biggl[\theta (b_{1}-b_{3})K_{0}\biggl(\sqrt{x_{3}(1-r_{2}^{2})}m_{B}b_{1}\biggl) \notag \\
&& \cdot I_{0}\biggl(\sqrt{x_{3}(1-r_{2}^{2})}m_{B}b_{3}\biggl)+\theta (b_{3}-b_{1})K_{0}\biggl(\sqrt{x_{3}(1-r_{2}^{2})}m_{B}b_{3}\biggl) \notag \\
&& \cdot I_{0}\biggl(\sqrt{x_{3}(1-r_{2}^{2})}m_{B}b_{1}\biggl)\biggl]S_{t}(x_{3}),
\\ 
h_{nfe}(x_{1},x_{2},x_{3},b_{1},b_{3}) = && \biggl\{\theta(b_{2}-b_{1}) I_{0}(m_{B}
\sqrt{x_{1}x_{3}(1-r_{2}^{2})}b_{1})K_{0}(m_{B}\sqrt{x_{1}x_{3}(1-r_{2}^{2})}b_{2})\hspace{0.5cm} \notag \\
&& + \ (b_{1}\leftrightarrow b_{2})\biggl\}\cdot \biggl(\begin{array}{cc} K_{0}(m_{B}F_{(1)}b_{2}),
& {\rm for} \;\;  F_{(1)}^{2}>0 \\ \frac{\pi i}{2}H_{0}^{(1)}(m_{B}\sqrt{|F_{(1)}|^{2}}b_{2}),
& {\rm for} \;\;  F_{(1)}^{2}<0
\end{array}
\biggl),
\eeq
where $J_{0}$ is the Bessel function, $K_{0}$ and $I_{0}$ are the modified Bessel functions with
$K_{0}(-ix)=-(\pi/2)Y_{0}(x)+i(\pi/2)J_{0}(x)$. The $F_{(1)}^{2}$ is defined by
\beq
F_{(1)}^{2}=(x_2-x_1)((x_3-x_2)r_2^2-x_3)+r_c^2.
\eeq
The expressions for the evolution functions $E_i(t)$ are defined as follows,
\beq
E_{fe}(t) &=& \alpha_s(t) \cdot \exp{[-S_{ab}(t)]}\cdot S_t(x) \;, \\
E_{nfe}(t) &=& \alpha_s(t) \cdot \exp{[-S_{cd}(t)]}\;,
\eeq
in which the factor $S_{t}(x)$ arising from threshold resummation is universal
and has been parameterized in a simplified form which is independent of the decay channels, twist, and flavors as~\cite{Li:2001ay, Li:2002mi}
\beq
S_{t}(x)=\frac{2^{1+2c}\Gamma(3/2+c)}{\sqrt{\pi}\Gamma(1+c)}[x(1-x)]^{c},
\eeq
with $c = 0.4$ and this factor is normalized to unity.
And the Sudakov factors $S_{ab}(t)$ and $S_{cd}(t)$ used in this paper are
given as the following,
\beq
S_{ab}(t)= &&s(x_{1}P_{1}^{+},b_{1})+ s(x_{3}P_{3}^{-},b_{3})+s((1-x_{3})P_{3}^{-},b_{3})
-\frac{1}{\beta_{1}}\biggl[\ln \frac{\ln(t/\Lambda)}{-\ln(b_{1}\Lambda)}+\ln \frac{\ln(t/\Lambda)}{-\ln(b_{3}\Lambda)}\biggl],
\\ 
S_{cd}(t)&=&s(x_{1}P_{1}^{+},b_{1})+ s_c(x_{2}P_{2}^{+},b_{2})+s_c((1-x_{2})P_{2}^{+},b_{2})
+ s(x_{3}P_{3}^{-},b_{1})+s((1-x_{3})P_{3}^{-},b_{1})
\non
&& \hspace{2cm} -\frac{1}{\beta_{1}}\biggl[2\ln \frac{\ln(t/\Lambda)}{-\ln(b_{1}\Lambda)}+\ln \frac{\ln(t/\Lambda)}{-\ln(m_c\Lambda)}\biggl]\;,
\eeq
where the functions $s(Q,b)$ and $s_c(Q,b)$ could be found easily in
Refs.~\cite{Keum:2000ph,Lu:2000em,Liu:2018kuo,Liu:2023kxr}.
And the running hard scale $t_{i}^{,}s$ in the above equations are chosen as
the maximum energy scale to kill the large logarithmic radiative corrections
and they are given as follows,
\beq
t_{a} &=& \max(\sqrt{x_{3}(1-r_{2}^{2})}m_{B},1/b_{1},1/b_{3}), \notag \\
t_{b} &=& \max(\sqrt{x_{1}(1-r_{2}^{2})}m_{B},1/b_{1},1/b_{3}), \notag \\
t_{nfe} &=& \max(\sqrt{x_{1}x_{3}(1-r_{2}^{2})}m_{B},\sqrt{\left|(x_2-x_1)[(x_3-x_2)r_2^2-x_3]+r_c^2\right|}
\ m_{B},1/b_{1},1/b_{2}).
\eeq

\end{appendix}

%
%

\end{document}